\documentclass[aps,prb,twocolumn,groupedaddress,amsmath,amssymb,amsfonts]{revtex4}

\usepackage{latexsym,amssymb}
\usepackage{graphicx,color,pstricks}
\usepackage{epsfig,epsf,bm}

\definecolor{gray}{rgb}{0.7,0.7,0.7}

\begin{document}

\title{Magnetic impurity in a triple-component semimetal}

\author{Yu-Li Lee}
\email{yllee@cc.ncue.edu.tw}
\affiliation{Department of Physics, National Changhua University of Education, Changhua, Taiwan, R.O.C.}

\author{Yu-Wen Lee}
\email{ywlee@thu.edu.tw}
\affiliation{Department of Applied Physics, Tunghai University, Taichung, Taiwan, R.O.C.}

\date{\today}

\begin{abstract}
 We investigate the effects of a magnetic impurity in a multiband touching fermion system, specifically,
 a triple-component semimetal with a flat band, which can be realized in a family of transition metal
 silicides (CoSi family). When the chemical potential coincides with the flat band, it is expected that
 the impurity response of this system will be very different from that of an ordinary Dirac or Weyl
 semimetal of which the density of states at the Fermi level vanishes. We first determine the phase
 diagram within the mean-field approximation. Then, we study the local moment regime by employing two
 different methods. In the low temperature regime, the Kondo screening is analyzed by the variational
 wavefunction approach and the impurity contributions to the magnetic susceptibility and heat capacity
 are obtained, while at higher temperature, we use the equation of motion approach to calculate the
 occupation number of the impurity site and the impurity magnetic susceptibility. The results are
 compared and contrasted with those in the usual Fermi liquid and the Dirac/Weyl semimetals.
\end{abstract}

\maketitle

\section{Introduction}

Weyl semimetals (WSMs) are topological metals that show interesting physics such as Fermi arc surface
states and chiral anomaly\cite{XWan}, and they have received a lot of attention recently. Their existence
in condensed matter systems has been predicted by theories\cite{XWan} and subsequently observed in
experiments\cite{CShekhar,BQLv,SYXu,BLv,LYang,SYXu2,NXu}. In these materials, the conduction and valence
bands touch each other at a few isolated points in the first Brillouin zone, know as the Weyl nodes. Near
the Weyl nodes, the WSM can be described by a pseudospin-$1/2$ which describes the valence and conduction
band degrees of freedom. Moreover, the Weyl nodes act as the sources or sinks of the Abelian Berry
curvature, so that they carry quantized magnetic monopole charges $Q=\pm 1$ in the momentum space. In the
presence of both the time-reversal and inversion symmetries, the bands are degenerate so that each node
consists of two Weyl fermions, resulting in the Dirac semimetal (DSM). Since the density of states (DOS)
of a DSM/WSM vanishes at the node, the thermodynamic response functions are quite different from those of
a usual Fermi liquid (FL) when the Fermi level coincides with the nodes.

One of the examples which distinguishes the FL and the DSM/WSM is the Kondo effect\cite{Kondo, Hewson}.
The Kondo physics results from the spin-flip scattering between the conduction electrons and a local
magnetic impurity, and has been studied by various different methods. In the FL, a spin-$1/2$ impurity
is completely screened at long distance by the conduction electrons. In the DSM/WSM, due to the special
property of the DOS mentioned above, when the chemical potential coincides with the nodes, the associated
magnetic impurity problem falls into the category of the pseudogap Kondo
problem\cite{Withoff, Cassanello,Gonzale,Vojta,Fritz}. That is, the Kondo screening in a system with
vanishing DOS at the Fermi level occurs only when the (antiferromagnetic) exchange coupling between
the conduction electrons and the impurity moment exceeds some critical strength. More recently, various
aspects of the problem of a magnetic impurity in a DSM or WSM, such as the scaling of the Kondo
temperature with respect to the doping\cite{Principi}, the interplay of long-range scalar disorder and
Kondo screening\cite{Principi}, the effect of various symmetry-breaking perturbations\cite{Mitchell},
and the spin-spin correlation function between the impurity spin and that of the conduction
electrons\cite{JHSun}, were further studied.

In the condensed matter system, it is possible to have more bands touching at a single point in the
Brillouin zone due to the crystal symmetry\cite{Manes,Bradlyn,PTang,GChang}. For a multiband touching
fermion system with $2J+1$ bands touching at a single point, where $J$ can be a half-integer or an
integer, it can be described in terms of a pseudospin $J$ representation of the SU($2$) Lie algebra in
the close proximity to the band touching points\cite{Manes,Bradlyn,PTang,Ezawa1,Ezawa2}. The WSM
corresponds to the $J=1/2$ case. The energy spectrum of a system with integer $J$ displays, in addition
to the $2J$ branches with nonzero Fermi velocities, a completely flat band which arises from the trivial
eigenvalue of the pseudospin operator. Similar to the WSM, the Berry curvature of the band indexed by
$\alpha$, where $\alpha=-J,-J+1,\cdots,J$, describes a monopole carrying the monopole charge $Q=2\alpha$
in the momentum space\cite{Ezawa2}. Moreover, the chiral anomaly and anomalous Hall effect are also
present in these systems\cite{Ezawa2}. Based on {\it ab initio} calculations, it was suggested that
triple-component ($J=1$) fermion systems can be realized in a family of transition metal silicides (CoSi
family) when the spin-orbital coupling (SOC) is weak\cite{PTang,GChang}. Recently, the $J=1$ fermion was
observed in a transition-metal silicide CoSi by using the angle-resolved photoemission
spectroscopy (ARPES)\cite{exp}.

In the present work, we would like to analyze the Kondo physics of a single magnetic impurity in a $J=1$
fermion system. In the local moment regime (i.e., the parameter regime in which the impurity behaves
like a local moment), the coupling between the magnetic impurity and the conduction electrons can be
described by the Hamiltonian
\begin{equation}
 H_K=K\bm{S}\cdot\bm{\tau}(0) \ , \label{jfkh1}
\end{equation}
where $\bm{S}$ is the impurity spin located at the position $\bm{r}=0$, $\bm{\tau}(0)$ is the spin
density of conduction electrons at the impurity site, and the constant $K$ is the exchange coupling
between them. When the conduction electrons are treated as noninteracting particles and characterized by
the DOS $N(\epsilon)$, one may integrate out the excitations in the energy shell $[e^{-l}\Lambda,\Lambda]$
in terms of a simple generalization of the ``poor-man's-scaling" method introduced by Anderson\cite{Anderson},
where $\Lambda$ is the UV cutoff in energies and $0<l\ll 1$. The resulting renormalization of $K$ to the
one-loop order is then
\begin{equation}
 \delta K=\frac{K^2}{2\Lambda}I \ , \label{jfkrg1}
\end{equation}
where
\begin{eqnarray*}
 I= \! \int^{\Lambda}_{e^{-l}\Lambda} \! d\epsilon N(\epsilon)=l\Lambda N(\Lambda)+O(l^2) \ .
\end{eqnarray*}
For the triple-component fermion system, we have
\begin{equation}
 N(\epsilon)=A\delta(\epsilon)+C\epsilon^2 \ , \label{jfdos1}
\end{equation}
where $A=C\Lambda^3/3$ and $C$ is a positive constant, provided that we set the energy at the band
touching point to be zero. We notes that $A=0$ for the WSM. For both the triple-component fermion system
and the WSM, we find that $N(\Lambda)=C\Lambda^2$. Accordingly, one may naively indicate that the
magnetic impurity in the $J=1$ fermion system belongs to the pseudogap Kondo problem\cite{foot1}, as what
happens in the WSM. That is, there exists a critical value $K_c$ such that the Kondo screening occurs
only when $K>K_c$.

On the other hand, when the Fermi level lies at $\epsilon=0$, the $J=1$ fermion system should be more FL
like due to the presence of the flat band. (Here we ignore the electron-electron interactions.) If this
is indeed the case, we expect that when the chemical potential $\mu=0$, the low temperature properties
of the magnetic impurity will behave like those for the usual Kondo problem in a FL. Since in the coarse
graining procedure, we just integrate out the high-energy degrees of freedom and do not take into account
the flat band at all, the renormalization group (RG) analysis discussed above may be questionable. One
of our motivation in this work is therefore to resolve this paradox and to determine whether the Kondo
physics in the $J=1$ fermion system belongs to the class of the pseudogap Kondo problem or that of the
usual Kondo problem in a FL.

We model this problem in terms of the Anderson impurity model. We first determine its phase diagram
within a mean-field approximation (Fig. \ref{jfkondof16}), and indicate the parameter regime that is
associated with the local moment physics in which we are interested. In particular, since the Kondo
physics is most transparent in the strong coupling regime of the Anderson impurity model, our analysis
will focus on the infinite-$U$ Anderson model. To capture the role of the flat band, we utilize two
complementary methods to analyze the physical properties of the local moment regime. We use a variational
wavefunction approach\cite{Yosida,Varma} to study the parameter dependence of the ground-state properties,
such as the binding energy and the impurity contribution $\chi_{imp}$ to the magnetic susceptibility.
Such a method is non-perturbative in nature and has been proved to be useful in the study of the related
problem in a WSM\cite{JHSun}. We find that in contrast with the prediction of the perturbative RG or the
``poor-man's-scaling", the binding energy is always positive in the local moment regime, implying the
occurrence of the Kondo screening. This binding energy is identified as the Kondo temperature. We also
calculate the parameter dependence of the binding energy (or the Kondo temperature), as shown in Fig.
\ref{jfkondof1}.

Furthermore, following the idea of local Fermi liquid description of the usual Kondo effect in a Fermi
liquid\cite{Hewson}, we propose an effective Hamiltonian $H_{eff}$ [Eq. (\ref{jffl1})] describing the
physics at the temperature much below the Kondo temperature. By calculating the local electron occupation
at the impurity site and the impurity contribution to the magnetic susceptibility at $T=0$ in terms of
both the variational wavefunction and $H_{eff}$, we are able to relate the parameters in $H_{eff}$ with
those in the infinite-$U$ Anderson model. Equipped with this, we plot the impurity spectral density at
$T=0$ (Fig. \ref{jfflf11}) and calculate the impurity contribution $C_{imp}$ to the heat capacity at low
temperature. In contrast with the Kondo effect in an ordinary FL, the Kondo resonance in the $J=1$
fermions is split into two peaks due to the presence of the flat band.

Above the Kondo temperature, we employ the equation of motion (EOM) approach\cite{Varma,Meir} within the
Hartree-Fock approximation to calculate the impurity Green's function from which we can extract the
temperature dependence of the occupation number of electrons at the impurity site (Fig. \ref{jfkondof13})
and $\chi_{imp}$ (Fig. \ref{jfkondof14}). We also compare our results with the Kondo problem in the
ordinary FL and in the pure WSM.

The present paper is organized as follows. In Sec. \ref{model}, we discuss the various terms in the
Hamiltonian of the Anderson impurity model and determine its phase diagram within a mean-field
approximation. The properties of the local moment regime is analyzed in terms of the variational
wavefunction and EOM methods, which are presented in Sec. \ref{vwf} and \ref{eom}, respectively. Our
results are summarized and discussed in the last section. The details of the calculations are listed in
the appendix.

\section{The model}
\label{model}
\subsection{The Hamiltonian}

We use the Anderson impurity model to describe a single magnetic impurity in a $J=1$ fermion system. The
corresponding Hamiltonian consists of three terms: $H=H_C+H_D+H_V$, where $H_C$, $H_D$, and $H_V$
describe the non-interacting $J=1$ fermions, the impurity fermions, and the hybridization between them,
respectively.

The three-band touching is assured at high-symmetry points by some nonsymmorphic space group symmetries
in certain lattice models\cite{Manes,Bradlyn,PTang,GChang,Ezawa1}. Here we will focus on the CoSi family.
As analyzed in Refs. \onlinecite{PTang,GChang}, the three-band touching will occur at the center of the
first BZ (the $\Gamma$ point) in the absence of SOC. Thus, there are sixfold degeneracy at the $\Gamma$
point (including spin). In the presence of the SOC, this sixfold degeneracy is split into two crossing
points with twofold and fourfold degeneracy, respectively. In a recent experiment on CoSi\cite{exp},
three-band touching was observed at the $\Gamma$ point, which implies that the SOC is very weak in this
family of materials.

Based on the above observation, the three-band touching in CoSi arises from the orbital dynamics of
electrons. The role played by the electron spin is similar to that in graphene. Hence, we write $H_c$ in
the form
\begin{equation}
 H_c=\! \sum_{\bm{p},\sigma}\Psi^{\dagger}_{\bm{p}\sigma}[H(\bm{p})-\mu]\Psi_{\bm{p}\sigma} \ ,
 \label{jfkh2}
\end{equation}
where $\Psi_{\bm{p}\sigma}=[\Psi_{1\bm{p}\sigma},\Psi_{0\bm{p}\sigma},\Psi_{-1\bm{p}\sigma}]^t$,
$\alpha=1,0,-1$ is the band index, $\sigma=\pm 1$ correspond respectively to up- and down-spins, $\mu$
is the chemical potential, and the annihilation and creation operators of electrons,
$\Psi_{\alpha\bm{p}\sigma}$ and $\Psi^{\dagger}_{\alpha\bm{p}\sigma}$, satisfy the canonical
anticommutation relations.

Near the band touching point $\bm{p}_0$, $H(\bm{p})$ can be written as\cite{Manes,Bradlyn,PTang}
\begin{equation}
 H(\bm{p})=v\bm{k}\cdot\bm{J} \ , \label{jfh1}
\end{equation}
where $\bm{k}=\bm{p}-\bm{p}_0$, $v>0$ is the Fermi velocity, and $\bm{J}=(J_x,J_y,J_z)$ is the
pseudospin-$1$ operator (with $J=1$) which obeys the SU($2$) Lie algebra
$[J_a,J_b]=i\epsilon_{abc}J_c$ with $a,b,c,=x,y,z$. At low energies, the physics is dominated by the
excitations around the band touching point. Hence, we will change the notation and write
$c_{\alpha\bm{k}\sigma}=\Psi_{\alpha\bm{k}+\bm{p}_0\sigma}$. It is clear that $c_{\alpha\bm{k}\sigma}$,
$c^{\dagger}_{\alpha\bm{k}\sigma}$ still obey the canonical anticommutation relations.

To find the spectrum of $H(\bm{p})$, we write $\bm{k}=k\hat{\bm{k}}$ where $k=|\bm{k}|$ and
$\hat{\bm{k}}=(\sin{\theta}\cos{\phi},\sin{\theta}\sin{\phi},\cos{\theta})$. Then, $H(\bm{p})$ can be
diagonalized by the unitary matrix $U$ as
\begin{equation}
 U(\bm{k})H(\bm{p})U^{\dagger}(\bm{k})=vkJ_z \ , \label{jfh11}
\end{equation}
where
\begin{equation}
 U(\bm{k})=e^{i\theta J_y}e^{i\phi J_z} \ . \label{jfh12}
\end{equation}
Thus, the energy spectrum of $H(\bm{k})$ is given by
\begin{equation}
 E_{\alpha}(k)=\alpha vk \ . \label{jfh13}
\end{equation}
Notice that Eq. (\ref{jfh11}) amounts to the following identity
\begin{equation}
 U(\bm{k})\hat{\bm{k}}\cdot\bm{J}U^{\dagger}(\bm{k})=J_z \ , \label{jfkh25}
\end{equation}
In terms of Eq. (\ref{jfh13}), it is straightforward to show that the DOS of the $J=1$ fermions around
the band touching point is indeed given by Eq. (\ref{jfdos1}).

The single-impurity Hamiltonian $H_D$ is of the form
\begin{equation}
 H_D=\! \sum_{\sigma}(\epsilon_d-\mu-\sigma h)n_{d\sigma}+Un_{d\uparrow}n_{d\downarrow} \ ,
 \label{jfkh20}
\end{equation}
where $n_{d\sigma}=d^{\dagger}_{\sigma}d_{\sigma}$ is the number operator of the impurity fermions with
spin $\sigma$, $\epsilon_d$ is the on-site energy of the impurity fermions, and $d_{\sigma}$ and
$d^{\dagger}_{\sigma}$ satisfy the canonical anticommutation relations. $h$ is the applied magnetic field.
Since we are only interested in the impurity contribution $\chi_{imp}$ to the magnetic susceptibility, we
consider only the coupling between $h$ and the impurity fermions. When the condition
\begin{equation}
 \epsilon_d<\mu<\epsilon_d+U \ , \label{jfkh21}
\end{equation}
is satisfied, the ground state of $H_D$ is singly occupied, i.e., a two-fold degenerate magnetic doublet.
When it is probed at energies much below the smallest charge excitation energy,
$\Delta E=\mbox{min}\{\mu-\epsilon_d,U+\epsilon_d-\mu\}$, only the spin degrees of freedom remain. Thus,
the impurity behaves like a local moment. This is the local moment regime in the atomic limit. In this
regime, the interaction between the local moment and conduction electrons is given by Eq. (\ref{jfkh1})
with $K>0$.

Finally, the hybridization between the $J=1$ fermions and impurity can be written as
\begin{equation}
 H_v=\frac{1}{\sqrt{\Omega}} \! \sum_{\bm{p},\sigma,\alpha} \! \left(
 V_{\bm{p}\alpha}\Psi^{\dagger}_{\alpha\bm{p}\sigma}d_{\sigma}+\mathrm{H.c.}\right) . \label{jfkh11}
\end{equation}
By expanding around the touching point $\bm{p}_0$, we may neglect the momentum dependence of the
hybridization amplitude, i.e., $V_{\bm{p}\alpha}\approx V_{\bm{p}_0\alpha}$. Moreover, for simplicity,
we assume that the impurity is equally coupled to the three bands, i.e., $V_{\bm{p_0}\alpha}=V$. In the
$U\rightarrow+\infty$ limit, we may obtain the Kondo coupling $K=|V|^2/|\epsilon_d-\mu|$ when Eq.
(\ref{jfkh21}) is satisfied and $|V|^2\ll|\epsilon_d-\mu|$\cite{Hewson}.

\subsection{The mean-field phase diagram}

In general, the Anderson impurity model has two regimes: the simple resonance and the local moment
regime\cite{Anderson2,Coleman}. The nature of the former is illustrated by the $U=0$ limit in which the
hybridization between conduction electrons and impurity fermions turns the impurity level into a virtual
bound state or resonance. On the other hand, the behavior of the magnetic impurity behaves like a free
moment at high temperatures in the local moment regime and the nature of that regime can be captured by
the $U\rightarrow+\infty$ limit.

On account of the competition between the hybridization and the on-site Coulomb repulsion $U$, the local
moment regime will be different from the one in the atomic limit. Since the Kondo effect occurs only in
the local moment regime, we need to determine the boundary between the local moment and simple resonance
regime before plunging into the detailed study.

To do it, we employ a mean-field decoupling of the $U$ term in $H_D$\cite{Anderson2}:
\begin{eqnarray*} 	
 Un_{d\uparrow}n_{d\downarrow}\rightarrow Un_{d\uparrow}\langle n_{d\downarrow}\rangle+U\langle
 n_{d\uparrow}\rangle n_{d\downarrow}-U\langle n_{d\uparrow}\rangle\langle n_{d\downarrow}\rangle \ .
\end{eqnarray*}
This results in the shift of the impurity level
\begin{eqnarray*}
 \epsilon_d\rightarrow\epsilon_{d\sigma}=\epsilon_d+U\langle n_{d-\sigma}\rangle \ .
\end{eqnarray*}
Hence, up to a constant term, the mean-field Hamiltonian is of the form
\begin{eqnarray*}
 H_{MF}=H_C+H_V+\! \sum_{\sigma}(\epsilon_{d\sigma}-\mu)n_{d\sigma} \ .
\end{eqnarray*}
Since $H_{MF}$ is quadratic in the fermion operators, it can be solved exactly.

The values of $\langle n_{d\sigma}\rangle$ at $\mu=0$ can be determined by the self-consistent equations
\begin{equation}
 \langle n_{d\sigma}\rangle=\! \int^0_{-\infty} \! \! d\omega\frac{\lambda\omega^4/D}
 {[L_{-\sigma}(\omega)]^2+(\pi\lambda\omega^3/D)^2} \ , \label{jfkmfe1}
\end{equation}
where
$L_{\sigma}(\omega)=(1+2\lambda)\omega^2-(\epsilon_d+U\langle n_{d\sigma}\rangle)\omega-s\lambda D^2$,
$\lambda=CD|V|^2$ is the dimensionless coupling between the $J=1$ fermions and the impurity, and $D$ is
the half band width. We express $\langle n_{d\sigma}\rangle$ by the total occupation number
$n_d=\langle n_{d\uparrow}\rangle+\langle n_{d\downarrow}\rangle$ and the local moment
$M=\langle n_{d\uparrow}\rangle-\langle n_{d\downarrow}\rangle$, and thus Eq. (\ref{jfkmfe1}) can be
written as
\begin{eqnarray}
 n_d \! \! &=& \! \! \! \sum_{\sigma} \! \int^0_{-\infty} \! \! d\omega\frac{\lambda\omega^4/D}
 {[L_{\sigma}(\omega)]^2+(\pi\lambda\omega^3/D)^2} \ , \label{jfkmfe11} \\
 M \! \! &=& \! \! - \! \sum_{\sigma}\sigma \! \int^0_{-\infty} \! \! d\omega\frac{\lambda\omega^4/D}
 {[L_{\sigma}(\omega)]^2+(\pi\lambda\omega^3/D)^2} \ , ~~~~\label{jfkmfe12}
\end{eqnarray}
where $L_{\sigma}(\omega)$ is now given by
$L_{\sigma}(\omega)=(1+2\lambda)\omega^2-[\epsilon_d+U(n_d+\sigma M)/2]\omega-s\lambda D^2$.

The local moment regime corresponds to $M\neq 0$. To determine the critical value of $U_c$ for given
$\lambda$, we set $M=0$ in Eq. (\ref{jfkmfe11}), yielding
\begin{equation}
n_d=\! \int^0_{-\infty} \! \! d\omega\frac{2\lambda\omega^4/D}{[L_c(\omega)]^2+(\pi\lambda\omega^3/D)^2}
\ , \label{jfkmfe13}
\end{equation}
where $L_c(\omega)=(1+2\lambda)\omega^2-(\epsilon_d+U_cn_d/2)\omega-s\lambda D^2$. On the other hand, we
expand the R.H.S. of Eq. (\ref{jfkmfe12}) to the linear order in $M$ and get
\begin{equation}
1=-\! \int^0_{-\infty} \! \! d\omega\frac{2\lambda\omega^5U_cL_c(\omega)/D}
{\{[L_c(\omega)]^2+(\pi\lambda\omega^3/D)^2\}^2} \ . \label{jfkmfe14}
\end{equation}
The values of $U_c$ and $n_d$ (at $U=U_c$) for given $\lambda$ can be obtained by solving Eqs.
(\ref{jfkmfe13}) and (\ref{jfkmfe14}).

\begin{figure}
\begin{center}
 \includegraphics[width=0.95\columnwidth]{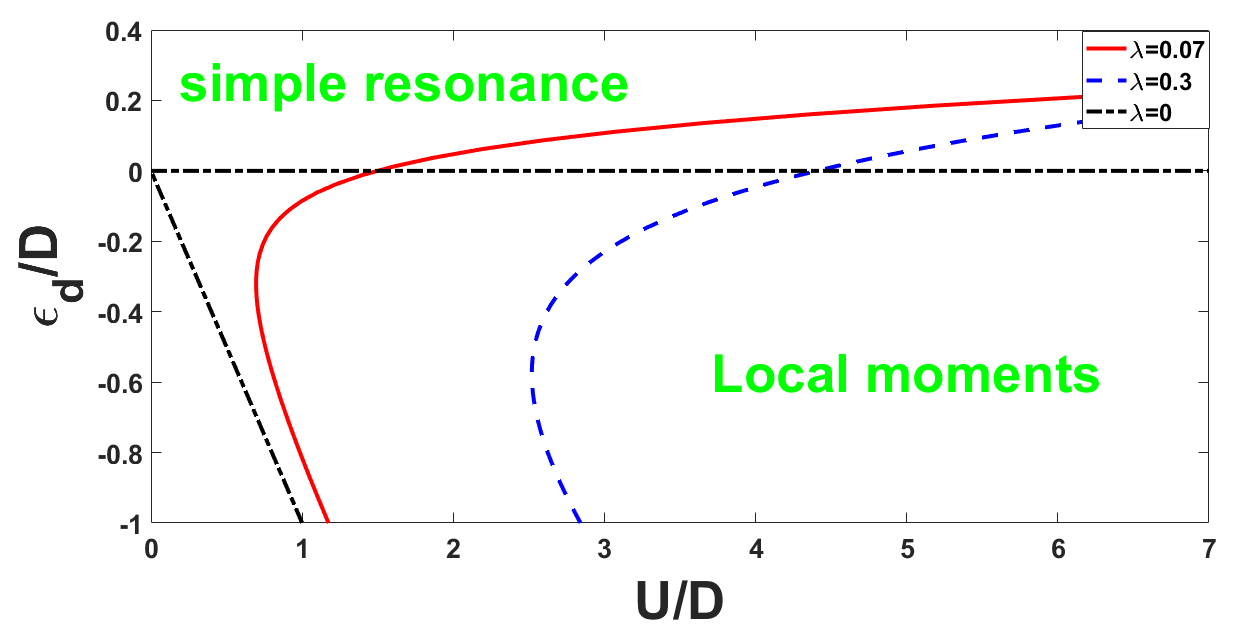}
 \caption{The mean-field phase diagram of the Anderson impurity model for the $J=1$ fermions at $\mu=0$
 with various values of the dimensionless hybridization strength $\lambda$. Note that $\lambda=0$
 corresponds to the atomic limit.}
 \label{jfkondof16}
\end{center}
\end{figure}

The resulting mean-field phase diagram is plotted in Fig. \ref{jfkondof16}. A few comments on it are in
order. First of all, in the presence of the hybridization, the local moment can appear when
$\epsilon_d>0$ for large enough values of $U$, in contrast with the case in a FL. This arises from the
fact that the single-peak structure in the spectral density becomes a two-peak structure and the
positions of the peaks are shifted away from $\epsilon_d$ due to the hybridization with the flat band.
(See Sec. \ref{resonance} for the details.) Next, when the value of $\lambda$ increases, the existence
of local moments requires a larger value of $U$ for given $\epsilon_d$. This must be the case since the
hybridization with the conduction electrons tends to screen the impurity level and turns it into a
resonance.

As we have mentioned in the introduction, since we are mainly interested in the Kondo physics which
reveals itself most clearly deep inside the local moment regime, we will concentrate on analyzing the
Anderson model in the infinite $U$ limit in the following.

\section{The variational wavefunction}
\label{vwf}

We now study the ground-state properties of $H$ in the $U\rightarrow+\infty$ limit with the help of the
variational wavefunction method. This is supposed to capture the essential properties of the whole local
moment regime.
To proceed, it is convenient to perform a unitary transformation at each $\bm{k}$ point
\begin{equation}
 \psi_{\bm{k}\sigma}=U(\bm{k})c_{\bm{k}\sigma} \ , ~~
 \psi^{\dagger}_{\bm{k}\sigma}=c^{\dagger}_{\bm{k}\sigma}U^{\dagger}(\bm{k}) \ , \label{jfkh22}
\end{equation}
where $U(\bm{k})$ is given by Eq. (\ref{jfh12}). With this transformation [Eq. (\ref{jfkh22})], $H_C$
takes the form
\begin{equation}
 H_C=\! \sum_{\bm{k},\sigma,\alpha}(\alpha vk-\mu)\psi^{\dagger}_{\alpha\bm{k}\sigma}
 \psi_{\alpha\bm{k}\sigma} \ , \label{jfkh23}
\end{equation}
while $H_V$ becomes
\begin{equation}
 H_V=\frac{1}{\sqrt{\Omega}} \! \sum_{\bm{k},\sigma,\alpha} \! \left(\tilde{V}_{\bm{k}\alpha}
\psi^{\dagger}_{\alpha\bm{k}\sigma}d_{\sigma}+\mathrm{H.c.}\right) , \label{jfkh24}
\end{equation}
where $\tilde{V}_{\bm{k}\alpha}=V \! \sum_{\beta}U_{\alpha\beta}(\bm{k})$.

\subsection{The binding energy}

In the absence of $H_V$, the ground state of $H_C$ is
\begin{equation}
 |\Psi_0\rangle=\prod_{\bm{k},\sigma,\alpha}^{\prime}\psi^{\dagger}_{\alpha\bm{k}\sigma}|0\rangle \ ,
 \label{jfkgs1}
\end{equation}
where $\prod^{\prime}$ means the product over states with energy below the Fermi level $\mu$. The
corresponding ground-state energy $E_0$ is of the form
\begin{equation}
 E_0=\epsilon_d-\mu-|h|+\! \sum_{\bm{k},\sigma,\alpha}^{\prime}(\alpha vk-\mu) \ . \label{jfkgs11}
\end{equation}
where $\sum^{\prime}$ means the sum over states with energy below the Fermi level.

\begin{figure}
	\begin{center}
		\includegraphics[width=0.95\columnwidth]{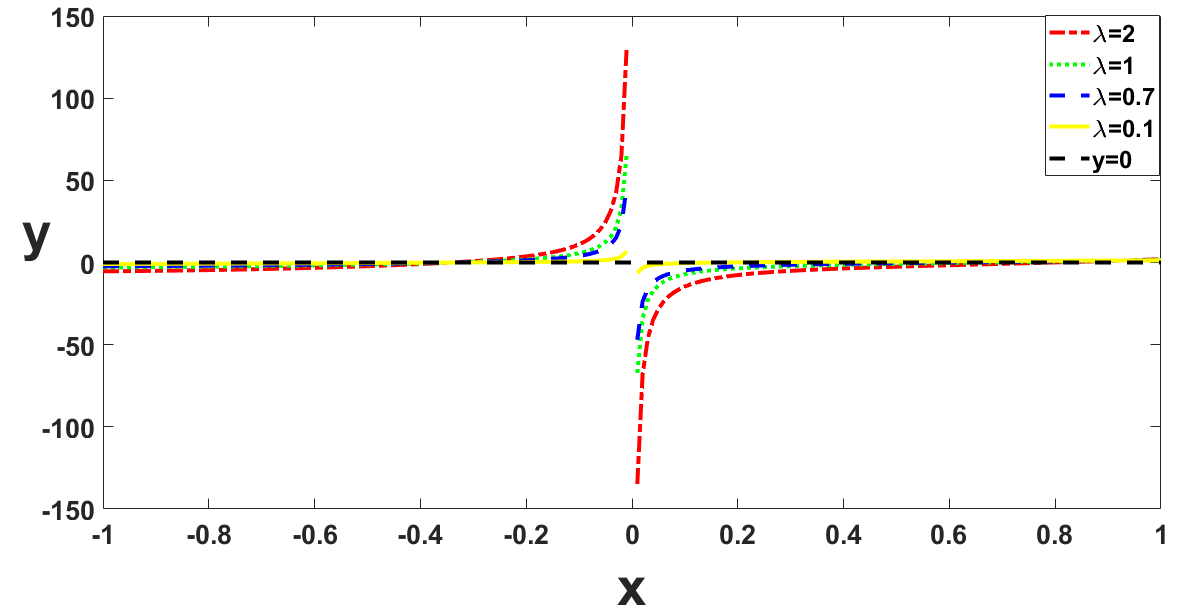}
		\caption{Graphic solution of Eq. (\ref{jfgse21}) with $\epsilon_d/D=-0.3$ for various values of
			$\lambda$, where $x=\Delta_b/D$ and $y=x+0.3-\lambda[2/(3x)+1-2x-2x^2\ln{|x|}]$.}
		\label{jfkondof11}
	\end{center}
\end{figure}

In the presence of $H_V$, motivated by the form of Eq. (\ref{jfkh24}), we try the ansatz for the ground
state\cite{JHSun, Varma}
\begin{equation}
 |\Psi\rangle=a_0|\Psi_0\rangle+\! \sum_{\bm{k},\sigma,\alpha}^{\prime}a_{\alpha\bm{k}\sigma}
 d^{\dagger}_{\sigma}\psi_{\alpha\bm{k}\sigma}|\Psi_0\rangle \ , \label{jfkgs12}
\end{equation}
in the $U\rightarrow+\infty$ limit. The variational energy $E$ for the trial state $|\Psi\rangle$ is
then given by
\begin{equation}
 E=\frac{\langle\Psi|H|\Psi\rangle}{\langle\Psi|\Psi\rangle} \ . \label{jfkgs13}
\end{equation}
According to the variational principle, the values of $a_0$ and $a_{\alpha\bm{k}\sigma}$ are determined
by the equations
\begin{eqnarray*}
 \frac{\partial E}{\partial a_0^*}=0=\frac{\partial E}{\partial a_{\alpha\bm{k}\sigma}^*} \ ,
\end{eqnarray*}
which lead to
\begin{eqnarray}
 & & \! \! \epsilon_d-\mu-|h|-\Delta_b=\frac{1}{\Omega} \! \sum_{\bm{k},\sigma,\alpha}^{\prime}
     \frac{|V|^2}{\alpha vk-\mu-\Delta_b-(\eta_h-\sigma)h} \nonumber \\
 & & \! \! =\! \sum_{\sigma} \! \int^{\mu}_{-D} \! \! d\epsilon\frac{|V|^2N(\epsilon)}
 {\epsilon-\mu-\Delta_b-(\eta_h-\sigma)h} \ , \label{jfgse15}
\end{eqnarray}
where $\Delta_b\equiv E_0-E$ is the binding energy and $\eta_h=\mbox{sgn}(h)$. The derivation of Eq.
(\ref{jfgse15}) is left in appendix \ref{a1}. Equation (\ref{jfgse15}) determines the value of $\Delta_b$
for given $|V|^2$. If $\Delta_b>0$, then the hybridized state $|\Psi\rangle$ has lower energy and is more
stable than the state $|\Psi_0\rangle$. This implies the occurrence of the Kondo effect.

\begin{figure}
	\begin{center}
		\includegraphics[width=0.95\columnwidth]{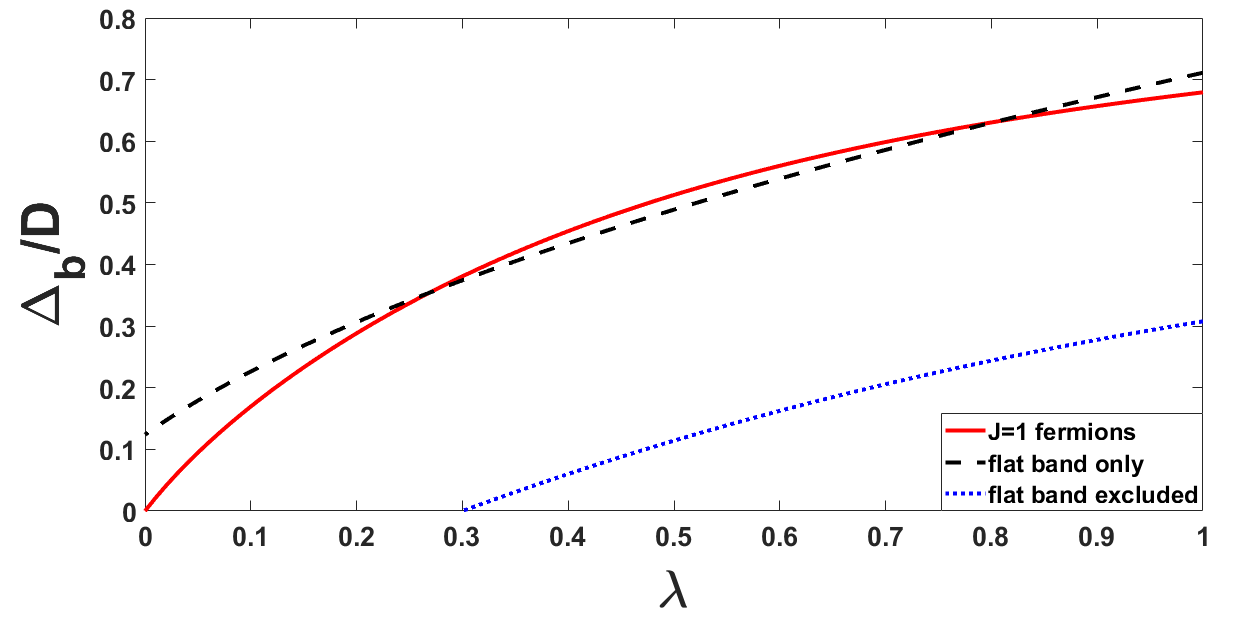}
		\includegraphics[width=0.95\columnwidth]{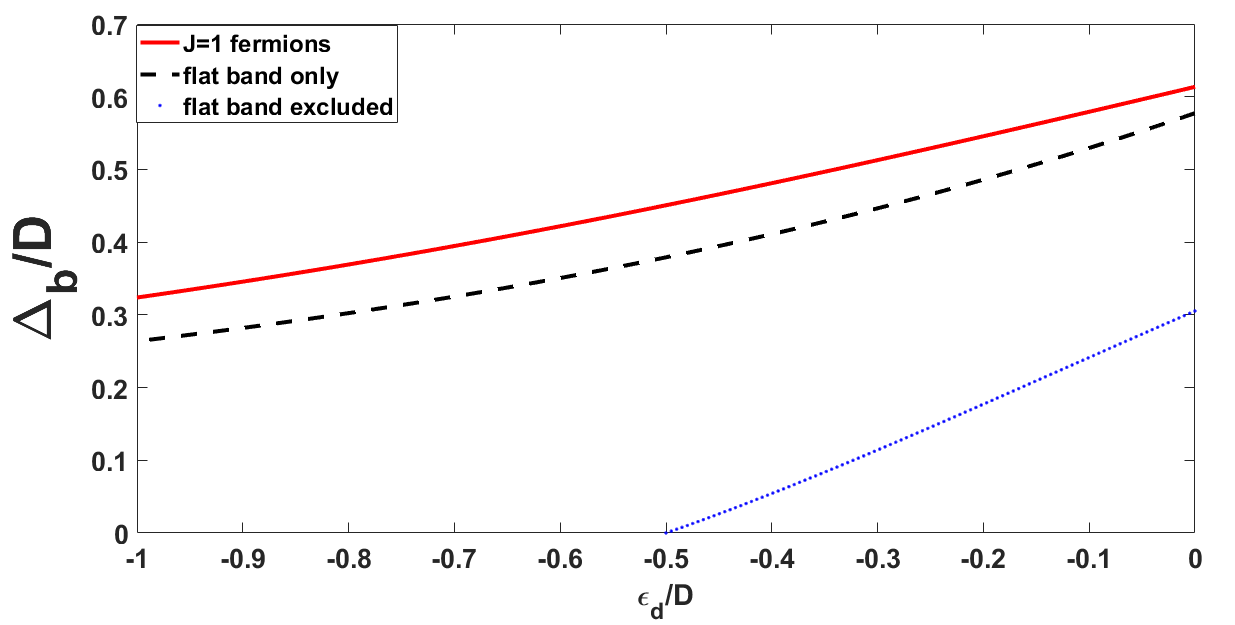}
		\caption{{\bf Top}: $\Delta_b/D$ as a function of $\lambda$ for the $J=1$ fermions (solid line), the
			case with the flat band only (dashed line), and the case in which the flat band is excluded (dotted line).
			In all situations, we take $\epsilon_d/D=-0.3$. {\bf Bottom}: $\Delta_b/D$ as a function of
			$\epsilon_d/D$ for the $J=1$ fermions (solid line), the case with the flat band only (dashed line), and
			the case in which the flat band is excluded (dotted line). In all situations, we take $\lambda=0.5$.}
		\label{jfkondof1}
	\end{center}
\end{figure}

We first consider the case with $h=0$. For this situation, Eq. (\ref{jfgse15}) reduces to
\begin{equation}
 \epsilon_d-\mu-\Delta_b=\! \int^{\mu}_{-D} \! \! d\epsilon\frac{2|V|^2N(\epsilon)}
 {\epsilon-\mu-\Delta_b} \ . \label{jfgse2}
\end{equation}
For $\mu=0^-$, the contribution from the flat band at $\epsilon=0$ vanishes and Eq. (\ref{jfgse2}) is
identical to the one for the WSM. On the other hand, for $\mu=0^+$, the contribution from the flat band
must be taken into account and Eq. (\ref{jfgse2}) becomes
\begin{equation}
 \frac{\Delta_b-\epsilon_d}{D}\approx\lambda \! \left(\frac{2D}{3\Delta_b}+1-\frac{2\Delta_b}{D}
 -\frac{2\Delta_b^2}{D^2}\ln{\! \left|\frac{\Delta_b}{D}\right|}\right) , \label{jfgse21}
\end{equation}
provided that $\mu,|\Delta_b|\ll D$. Figure \ref{jfkondof11} shows the graphic solution of Eq.
(\ref{jfgse21}) with $\epsilon_d/D=-0.3$. We see that there is always a solution with $\Delta_b>0$ as
long as $\epsilon_d<0$ and $\lambda>0$. This result is in contrast with the prediction of the
perturbative RG, which indicates the existence of a critical value of $\lambda$. This implies that the
latter cannot capture the physics of the flat band.

Figure \ref{jfkondof1} sketches $\Delta_b/D$ as a function of $\lambda$ with given $\epsilon_d/D$. The
same figure also exhibits the case with the flat band only and the one in which the flat band is
excluded. The last situation has solutions with $\Delta_b>0$ only when $\lambda>\lambda_c$. In all
cases, $\Delta_b/D$ is an increasing function of $\lambda$ for given $\epsilon_d/D<0$.

We also plot $\Delta_b/D$ as a function of $\epsilon_d/D$ with given $\lambda$ in Fig. \ref{jfkondof1}.
The same figure also exhibits the case with the flat band only and the one in which the flat band is
excluded. The last situation has solutions with $\Delta_b>0$ only when $\lambda>\lambda_c$ for given
$\epsilon_d<0$ or $\epsilon_d>\epsilon_{dc}$ for given $\lambda$. In all cases, $\Delta_b/D$ is an
increasing function of $\epsilon_d/D$ for given $\lambda$. We notice that the qualitative behavior of
$\Delta_b/D$ v.s. $\epsilon_d/D$ is similar to that for the Kondo effect in a FL.

When $\lambda<0.1$, $\Delta_b/D\ll 1$ and we may neglect the last term at the R.H.S. of Eq.
(\ref{jfgse21}). This results in an analytic expression for the binding energy:
\begin{equation}
 \frac{\Delta_b}{D}=\frac{\lambda+\epsilon_d/D+\sqrt{(\lambda+\epsilon_d/D)^2+(8/3)\lambda(1+2\lambda)}}
 {2(1+2\lambda)} \ . \label{jfgse22}
\end{equation}
Equation (\ref{jfgse22}) holds as long as $0<\lambda<0.1$ and $\epsilon_d<0$, and we have checked it
numerically. From Eq. (\ref{jfgse22}), we find that
\begin{equation}
 \frac{\Delta_b}{D}\approx\frac{\lambda}{3} \! \left|\frac{D}{\epsilon_d}\right| , \label{jfgse27}
\end{equation}
as $\lambda\rightarrow 0^+$ for given $\epsilon_d<0$. That is, $\Delta_b$ approaches zero as a linear
function of $\lambda$. This behavior is in contrast with the usual FL. There, the binding energy is
given by
\begin{eqnarray*}
 \Delta_b=\epsilon_Fe^{-1/[2N(0)|V|^2/|\epsilon_d|]} \ ,
\end{eqnarray*}
for $\Delta_b\ll |\epsilon_d|$, where $N(0)$ is the DOS at the Fermi energy $\epsilon_F$ for a single
species of fermions.

In the Kondo problem, there is a characteristic energy scale -- the Kondo temperature $T_K$ (by setting
$k_B=1$). There are various ways to define it, which differ from each other by a constant of $O(1)$. For
example, in the perturbative RG approach, $T_K$ is defined as the energy scale at which the renormalized
coupling diverges\cite{Hewson}. In the large-$N$ mean field treatment, it is defined as the highest
temperature for which the self-consistent equations have a nontrivial solution\cite{Withoff,Cassanello,Principi}.
In the variational wavefunction approach, the bound state will disappear at the temperature of an order
of $\Delta_b$. Thus, we may define the Kondo temperature as the binding energy, i.e.,
$T_K=\Delta_b$\cite{Yosida,Varma,Grosso}.

\subsection{The impurity properties at $T=0$}

In terms of the above results, we first compute the impurity contribution $\chi_{imp}$ to the magnetic
susceptibility at $T=0$ when $\mu=0^+$. By choosing $h>0$, Eq. (\ref{jfgse15}) can be written as
\begin{widetext}
 \begin{equation}
 \frac{\Delta_b+h-\epsilon_d}{D}\approx\lambda \! \left[\frac{sD}{\Delta_b}+\frac{sD}{(\Delta_b+2h)}+1
 -\frac{2(\Delta_b+h)}{D}-\frac{\Delta_b^2}{D^2}\ln{\! \left|\frac{\Delta_b}{D}\right|} \!
 -\frac{(\Delta_b+2h)^2}{D^2}\ln{\! \left|\frac{\Delta_b+2h}{D}\right|}\right] , \label{jfgse23}
 \end{equation}
\end{widetext}
for $h,|\Delta_b|\ll D$, where $s=A/(CD^3)=1/3$ for the $J=1$ fermions and $s=0$ for the WSM. To
calculate $\chi_{imp}$, we write $\Delta_b$ as $\Delta_b=\Delta_0+c_1h+c_2h^2+O(h^3)$ where $\Delta_0$
is the value of $\Delta_b$ at $h=0$ and $c_1$, $c_2$ are constants independent of $h$. Inserting this
expansion into Eq. (\ref{jfgse23}), we find that $c_1=-1$ and
\begin{eqnarray*}
 Dc_2=\frac{sD/\Delta_0+(\Delta_0/D)^2[\ln{(D/\Delta_0)}-3/2]}
 {s+b(\Delta_0/D)^2-(\Delta_0/D)^3[2\ln{(D/\Delta_0)}-1]} \ ,
\end{eqnarray*}
where $b=1/(2\lambda)+1$. For $\Delta_0/D\ll 1$,
\begin{eqnarray*}
 c_2\approx \frac{1}{\Delta_0[1+(b/s)(\Delta_0/D)^2]} \ .
\end{eqnarray*}

The ground-state energy at small $h$ is given by
\begin{eqnarray*}
 E=E_0-\Delta_b=E_0-\Delta_0-c_1h-c_2h^2+O(h^3) \ .
\end{eqnarray*}
$\chi_{imp}$ is defined by
\begin{eqnarray*}
 \chi_{imp}=\! \left.-\frac{\partial^2E}{\partial h^2}\right|_{h=0} \! =2c_2 \ .
\end{eqnarray*}
As a result, we get
\begin{equation}
 \chi_{imp}=\frac{2}{\Delta_0[1+(b/s)(\Delta_0/D)^2]} \ . \label{jfgse24}
\end{equation}
If the flat band is excluded by setting $s=0$, $c_2$ would become
\begin{eqnarray*}
 c_2=\frac{\ln{(D/\Delta_0)}-3/2}{D\{b-(\Delta_0/D)[2\ln{(D/\Delta_0)}-1]\}} \ ,
\end{eqnarray*}
leading to
\begin{eqnarray*}
 \chi_{imp}=\frac{2\ln{(D/\Delta_0)}-3}{D\{b-(\Delta_0/D)[2\ln{(D/\Delta_0)}-1]\}}\sim\frac{1}{D} \ ,
\end{eqnarray*}
which is vanishingly small. That is, the main contribution to $\chi_{imp}$ at $T=0$ arises from the
hybridization of the flat band and the impurity.

Next, we would like to compute the $d$-level occupation number $\langle n_d\rangle_0$ at $T=0$. By
definition, we have
\begin{eqnarray*}
 \langle n_d\rangle_0=\frac{\langle\Psi|n_d|\Psi\rangle}{\langle\Psi|\Psi\rangle}
 =\frac{\! \sum_{\bm{k},\alpha,\sigma}^{\prime}|a_{\alpha\bm{k}\sigma}|^2}{\langle\Psi|\Psi\rangle}
 =1-\frac{|a_0|^2}{\langle\Psi|\Psi\rangle} \ .
\end{eqnarray*}
Since $a_0\neq 0$, $\langle n_d\rangle_0<1$. In terms of Eq. (\ref{jfgse11}), we find that
\begin{eqnarray*}
 & & \! \! \langle\Psi|\Psi\rangle=|a_0|^2 \! \left[1+\frac{1}{\Omega} \!
	 \sum_{\bm{k},\alpha,\sigma}^{\prime}\frac{|\tilde{V}_{\bm{k}\alpha}|^2}{(\alpha vk-\Delta_b)^2}
	 \right] \\
 & & \! \! =|a_0|^2 \! \left\{1+\frac{2|V|^2}{\Omega} \! \sum_{\bm{k},\alpha,\beta}^{\prime} \! \left[
	 U^{\dagger}(\bm{k})[G(\Delta_b,\bm{k})]^2U(\bm{k})\right]_{\alpha\beta} \! \right\} \\
 & & \! \! =|a_0|^2 \! \left(1+\frac{2\lambda D^2}{3\Delta_b^2}\right) ,
\end{eqnarray*}
when $\mu\rightarrow 0^+$.

Accordingly, the $d$-level occupation number at $T=0$ is
\begin{equation}
 1-\langle n_d\rangle_0=\frac{1}{1+\frac{2\lambda D^2}{3\Delta_b^2}} \ . \label{jfgse28}
\end{equation}
Following from Eq. (\ref{jfgse27}),
\begin{eqnarray*}
 \frac{\Delta_b^2}{\lambda D^2}\approx\frac{\lambda}{9} \! \left(\frac{D}{\epsilon_d}\right)^{\! 2} \!
 \ll 1 \ ,
\end{eqnarray*}
for $\lambda\ll 0.1$, we conclude that $1-\langle n_d\rangle_0\ll 1$ when $\lambda\ll 0.1$.

\subsection{Local Fermi liquid theory}

When $T\ll T_K$, the impurity spin is completely shielded, as suggested by the variational wavefunction
approach. The impurity fermion and the conduction electrons within the shell, of width about $T_K$,
around the Fermi surface form a spin singlet. As a result, the impurity is no longer magnetic. It acts
as a structureless scatterer. Although the impurity degree of freedom disappears from this problem at
low temperatures, the effective Hamiltonian does not reduce to the pure potential scattering. This
singlet displays polarizability, which provides an indirect interaction between electrons located in the
vicinity of this singlet: one of the electrons polarizes the singlet, this polarization acts on another
electron. Thus, a local interaction arises between the electrons.

Based on this picture, we propose that the effective Hamiltonian close to the Kondo fixed point can be
written as\cite{Hewson}
\begin{eqnarray}
 H_{eff} \! \! &=& \! \! H_c+\frac{1}{\sqrt{\Omega}} \! \sum_{\bm{k},\alpha,\sigma} \! \left(T_{\bm{k}}
 c^{\dagger}_{\alpha\bm{k}\sigma}f_{\sigma}+\mathrm{H.c.}\right) \nonumber \\
 \! \! & & \! \! +(\epsilon_0-\mu)n_f+\tilde{U}\delta n_{f\uparrow}\delta n_{f\downarrow} \ ,
 \label{jffl1}
\end{eqnarray}
where $n_{f\sigma}=f^{\dagger}_{\sigma}f_{\sigma}$, $n_f=n_{f\uparrow}+n_{f\downarrow}$,
$\delta n_{f\sigma}=n_{f\sigma}-\langle n_{f\sigma}\rangle_0$, and $\langle\cdots\rangle_0$ is the
ground-state expectation value. Here the scatterer formed by the singlet is modeled by a virtual level,
of energy $\epsilon_0$, hybridized with the conduction electrons, with the hybridization strength
$T_{\bm{k}}$. The operators $f_{\sigma}$ and $f^{\dagger}_{\sigma}$, which obey the canonical
anticommutation relations, describe the electrons around the scatterer and the $\tilde{U}$ term is the
local interaction between electrons induced by the singlet. (Here we have treated the singlet as a point
scatterer.)

The fixed-point Hamiltonian is given by $H_{eff}[\tilde{U}=0]$, while the $\tilde{U}$ term is the leading
irrelevant operator. Following the spirit of Landau's FL theory, the effects of the $\tilde{U}$ term can
be obtained by the Hartree-Fock approximation or the perturbative expansion in $\tilde{U}$. Notice that
$\langle\delta n_{f\sigma}\rangle\neq 0$ may arise from thermal fluctuations at finite temperature or
the presence of applied fields. Hence, the $\tilde{U}$ term may be neglected when its contribution is
subleading compared to the one from the fixed-point Hamiltonian $H_{eff}[\tilde{U}=0]$ in the zero
temperature or zero field limits.

To relate the parameters in $H_{eff}$ to those in the infinite-$U$ Anderson model, we can employ
$H_{eff}$ [Eq. (\ref{jffl1})] to calculate some physical quantities and identify them as those obtained
from the variational wavefunction approach. For simplicity, we will set $T_{\bm{k}}=T_0$. Here we shall
choose $\langle n_f\rangle_0$ and $\chi_{imp}^{(0)}$ at $T=0$, where $\chi_{imp}^{(0)}$ denotes the
impurity contribution to the magnetic susceptibility given by the fixed-point Hamiltonian
$H_{eff}[\tilde{U}=0]$. Both can be obtained from Eqs. (\ref{jfchi1}) and (\ref{jfchi11}).

\begin{figure}
	\begin{center}
		\includegraphics[width=0.95\columnwidth]{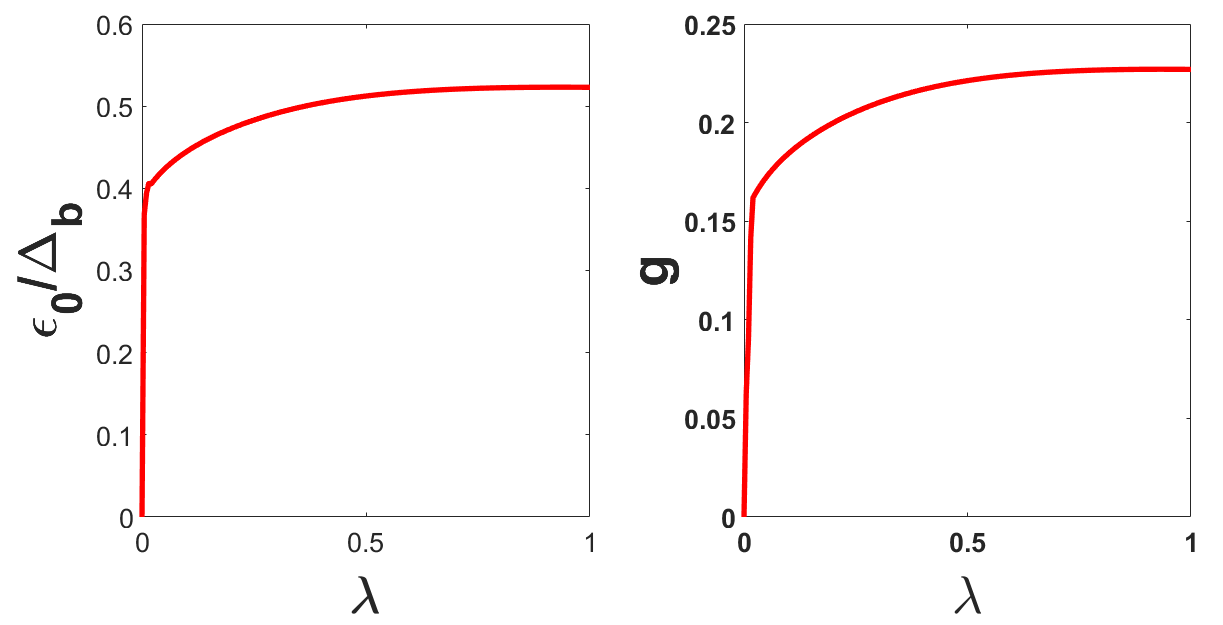}
		\caption{$\epsilon_0/\Delta_b$ and $g$ as functions of $\lambda$ with $\epsilon_d/D=-0.3$.}
		\label{jfflf1}
	\end{center}
\end{figure}

We will identify $\langle n_f\rangle_0$ as $\langle n_d\rangle_0$ given by Eq. (\ref{jfgse28}) and
$\chi_{imp}^{(0)}$ as $\chi_{imp}$ given by Eq. (\ref{jfgse24}). Then, we get
\begin{equation}
 \! \int_0^{+\infty} \! \! dx\frac{2gx^4}{[\tilde{L}_0(x)]^2+(\pi gx^3)^2}=\frac{1}
 {1+3(\Delta_b/D)^2/(2\lambda)} \ , \label{jffl12}
\end{equation}
and
\begin{equation}
 \! \int_0^{+\infty} \! \! dx\frac{4gx^5\tilde{L}_0(x)}{\{[\tilde{L}_0(x)]^2+(\pi gx^3)^2\}^2}=\frac{1}
 {1+(b/s)(\Delta_b/D)^2} \ , \label{jffl13}
\end{equation}
where $\tilde{L}_0(x)=(1+2g)x^2+(\epsilon_0/\Delta_b)x-g/3$ and $g=C\Delta_b|T_0|^2$ is the
(dimensionless) renormalized hybridization strength. The UV cutoff in energy in $H_{eff}$ is of
$O(\Delta_b)$. Different choices of it will lead to different values of $\epsilon_0$ and $g$ for given
$\lambda$ and $\epsilon_d/D$. Without loss of generality, we take it to be $\Delta_b$.

Equations (\ref{jffl12}) and (\ref{jffl13}) determine the fixed-point values of $g$ and
$\epsilon_0/\Delta_b$ for given $\lambda$ and $\epsilon_d/D$. Figure \ref{jfflf1} shows
$\epsilon_0/\Delta_b$ and $g$ as functions of $\lambda$ with a given value of $\epsilon_d/D$. Both
$\epsilon_0/\Delta_b$ and $g$ are increasing functions of $\lambda$ for a given value of $\epsilon_d/D$.
In particular, $\epsilon_0/\Delta_b$ and $g$ both increase rapidly with increasing $\lambda$ for
$\lambda<\lambda_*$, and their values increase slowly when $\lambda>\lambda_*$. In general, $\lambda_*$
is a function of $\epsilon_d/D$ and $\lambda_*\approx 0.05$ for $\epsilon_d/D=-0.3$.

The Kondo limit is usually defined by $\langle n_d\rangle_0=1$. This is achieved when
\begin{eqnarray*}
 \frac{1}{\sqrt{\lambda}} \! \left(\frac{\Delta_b}{D}\right) \! \rightarrow 0 \ ,
\end{eqnarray*}
following from Eq. (\ref{jfgse28}). In this limit, $\chi_{imp}=2/\Delta_b$ at $T=0$. Further, we find
from Eqs. (\ref{jffl12}) and (\ref{jffl13}) that $\frac{\epsilon_0}{\Delta_b}=0.2537$ and $g=0.2584$ in
the Kondo limit.

\begin{figure}
	\begin{center}
		\includegraphics[width=0.95\columnwidth]{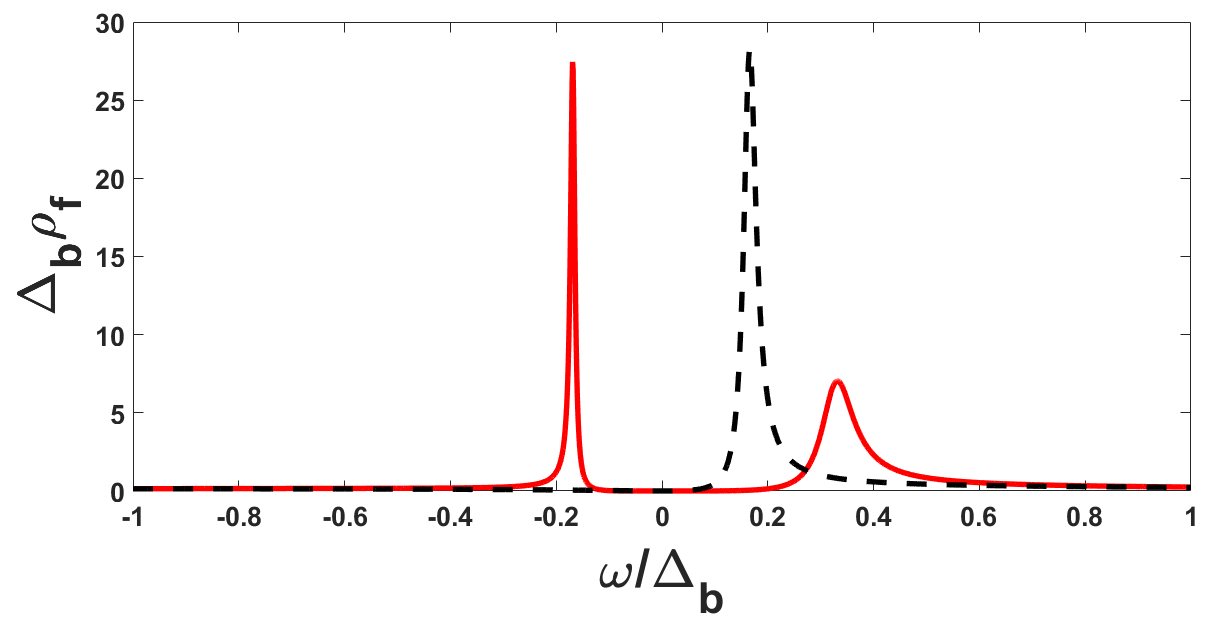}
		\caption{The spectral density at the impurity site, as a function of $\omega$ (in units of $\Delta_b$),
			in the Kondo limit (solid line). The dashed line shows the spectral density at the impurity site in the
			Kondo limit by removing the flat band.}
		\label{jfflf11}
	\end{center}
\end{figure}

With the help of Eq. (\ref{jfkgf12}), the spectral density at the impurity site is given by
\begin{equation}
 \rho_f(\omega)=\frac{(2g\omega^4/\Delta_b)}{[L_0(\omega)]^2+(\pi g\omega^3/\Delta_b)^2} \ ,
 \label{jffl15}
\end{equation}
at $T=0$, where $L_0(\omega)=(1+2g)\omega^2-\epsilon_0\omega-g\Delta_b^2/3$. We plot $\Delta_b\rho_f$ as
a function of $\omega/\Delta_b$ in the Kondo limit in Fig. \ref{jfflf11}. The qualitative feature remains
intact for other parameter values. We see that $\rho_f(\omega)$ exhibits two peaks close to the Fermi
energy ($\mu=0$ in the present case), which is different from the usual FL. For the latter, $\rho_f$ has
a single peak slightly above the Fermi energy and exactly at the Fermi energy in the Kondo limit), which
is known as the Kondo resonance. If we artificially remove the flat band, the spectral density will
exhibit a single peak slightly above the Fermi energy. This indicates that the splitting of the Kondo
resonance in the $J=1$ fermions arises from the flat band. On the other hand, the widths of both peaks
are about $0.1\Delta_b$ or smaller. This is similar to the usual FL.

Finally, we may employ $H_{eff}$ to calculate the impurity corrections to the thermodynamic response
functions at $T\ll T_K$. By integrating out the conduction electrons, the partition function with
$\tilde{U}=0$ can be written as
\begin{eqnarray*}
 Z=Z_0 \! \int \! \! D[f_{\sigma}]D[f^{\dagger}_{\sigma}]
 exp{\! \left[-\! \int^{\beta}_0 \! \! d\tau \! \sum_{\sigma}f^{\dagger}_{\sigma}[-D^{-1}_{\sigma}(\tau)]f_{\sigma}\right]}
 ,
\end{eqnarray*}
where $Z_0$ is the partition function of bulk electrons in the absence of impurity and the Fourier
transform of $D_{\sigma}(\tau)$ is given by
\begin{eqnarray*}
 \tilde{D}_{\sigma}(i\omega_n)=\frac{i\omega_n}
 {L_{\sigma}(i\omega_n)+i\pi\mbox{sgn}(\omega_n)g(i\omega_n)^3/\Delta_b} \ ,
\end{eqnarray*}
where $L_{\sigma}(\omega)=(1+2g)\omega^2-\epsilon_{\sigma}\omega-g\Delta_b^2/3$ and
$\epsilon_{\sigma}=\epsilon_0-\sigma h$. (This can be found with a procedure similar to the one to get
Eq. (\ref{jfkgf10}).) Consequently, the free energy can be written as $F=F_0+F_{imp}$ where $F_0$ is the
free energy of bulk electrons in the absence of the impurity and
\begin{eqnarray*}
 F_{imp} \! \! &=& \! \! -\frac{1}{\beta} \! \sum_{n,\sigma}
 \ln{\! \left[-\frac{L_{\sigma}(i\omega_n)+i\pi\mbox{sgn}(\omega_n)g(i\omega_n)^3/\Delta_b}{i\omega_n}\right]}
 \\
 \! \! & & \! \! \times e^{i\omega_n0^+} \ .
\end{eqnarray*}
The frequency summation can be performed with the help of contour integration, yielding
\begin{equation}
 F_{imp}=\! \sum_{\sigma} \! \int^{+\infty}_{-\infty} \! \frac{d\omega}{\pi}f(\omega)
 \tan^{-1}{\! \left[\frac{\pi g\omega^3/\Delta_b}{L_{\sigma}(\omega)}\right]} , \label{kondojffe1}
\end{equation}
where $f(\omega)=1/(e^{\beta\omega}+1)$ is the Fermi-Dirac distribution.

To compute the impurity correction $C_{imp}$ to the heat capacity, we set $h=0$ in $F_{imp}$. Since we
are interested only in the behavior of $C_{imp}$ at $T\ll T_K$, it suffices to expand $F_{imp}$ in powers
of $T/\Delta_b$. This can be achieved with the help of the standard Sommerfeld expansion, and we find
that
\begin{eqnarray*}
 F_{imp}=G(0)-\frac{7\pi^5}{20\Delta_b^3}T^4+O(T^6) \ ,
\end{eqnarray*}
where
\begin{eqnarray*}
 G(\epsilon)\equiv \! \int^{\epsilon}_{-\infty} \! \! d\omega
 \tan^{-1}{\! \left[\frac{\pi g\omega^3/\Delta_b}{L_0(\omega)}\right]} .
\end{eqnarray*}
As a result, $C_{imp}$ is of the form
\begin{equation}
 C_{imp}(T)=\frac{21\pi^5}{5} \! \left(\frac{T}{\Delta_b}\right)^{\! 3} \! +O(T^5) \ , \label{kondojffe11}
\end{equation}
when $T\ll\Delta_b$. Curiously, the prefactor in the leading term is universal, irrespective of
$\epsilon_0$ and $g$.

A few comments on the above results are in order. First of all, the correction arising from the
$\tilde{U}$ term is subleading because $\langle\delta n_{f\sigma}\rangle\rightarrow 0$ as $T\rightarrow 0$.
Next, let us compare $C_{imp}$ with the heat capacity per unit volume $c_0(T)$ for a non-interacting $J=1$
Fermi gas. The latter is given by
\begin{equation}
 c_0(T)=\frac{14\pi^4}{15}CT^3 \ . \label{kondojffe12}
\end{equation}
$c_0(T)$ is completely arises from the topologically nontrivial bands and the flat band does not
contribute to $c_0$ at all due to the lack of nontrivial dispersion. We see that $C_{imp}$ has the same
temperature dependence as $c_0$ when $T\ll T_K$.

\section{The equation of motion}
\label{eom}

To study the physical properties in the local moment regime at high temperature, i.e., for $T>T_K$, we
would like to calculate the single-particle Green function of the impurity fermions, which in the
imaginary-time formulation is defined as
\begin{equation}
 D_{\sigma\sigma^{\prime}}(\tau)\equiv -\langle\mathcal{T}_{\tau}\{d_{\sigma}(\tau)
 d^{\dagger}_{\sigma^{\prime}}(0)\}\rangle \ . \label{jfkgf1}
\end{equation}
Following Refs. \onlinecite{Varma} and \onlinecite{Meir}, we will calculate $D_{\sigma\sigma^{\prime}}$
in terms of its EOM.

In terms of the standard procedure, $D_{\sigma\sigma^{\prime}}(\tau)$ satisfies the equation
\begin{eqnarray}
 & & [i\omega_n-\epsilon_d+\mu+\sigma h-\Sigma_0(i\omega_n)]\tilde{D}_{\sigma\sigma^{\prime}}(i\omega_n)
     \nonumber \\
 & & =\delta_{\sigma\sigma^{\prime}}+U\tilde{C}_{\sigma\sigma^{\prime}}(i\omega_n) \ , \label{jfeom15}
\end{eqnarray}
where $\omega_n=(2n+1)\pi T$, $\tilde{A}(i\omega_n)$ is the Fourier transform of $A(\tau)$,
\begin{equation}
 C_{\sigma\sigma^{\prime}}(\tau)\equiv-\langle\mathcal{T}_{\tau}\{n_{d-\sigma}d_{\sigma}(\tau)
 d^{\dagger}_{\sigma^{\prime}}(0)\}\rangle \ , \label{jfkgf11}
\end{equation}
and
\begin{eqnarray*}
 & & \Sigma_0(i\omega_n)=\frac{1}{\Omega} \! \sum_{\bm{k},\alpha}\frac{|\tilde{V}_{\bm{k}\alpha}|^2}
	 {i\omega_n-\alpha vk+\mu} \\
 & & =\frac{1}{\Omega} \! \sum_{\bm{k},\alpha}\frac{|V|^2}{i\omega_n-\alpha vk+\mu}
	 =\! \int^{+\infty}_{-\infty} \! \! d\epsilon\frac{|V|^2N(\epsilon)}{i\omega_n-\epsilon+\mu} \ ,
\end{eqnarray*}
is the self-energy of the impurity fermions. In the following, we will focus on the $\mu=0$ case.

To illustrate the physics in the local moment regime and connect it to the previous variational
wavefunction approach, we consider the $U\rightarrow +\infty$ limit. When $U\neq 0$, an exact solution
of $\tilde{D}_{\sigma\sigma^{\prime}}(i\omega_n)$ cannot be obtained. To proceed, we need to make some
approximation. Within the Hartree-Fock approximation approximation\cite{Varma,Meir}, i.e.,
\begin{eqnarray}
 & & \langle\mathcal{T}_{\tau}\{\psi^{\dagger}_{\alpha\bm{k}-\sigma}d_{-\sigma}d_{\sigma}(\tau)
     d^{\dagger}_{\sigma}(0)\}\rangle\rightarrow 0 \nonumber \\
 & & \langle\mathcal{T}_{\tau}\{\psi_{\alpha\bm{k}-\sigma}d^{\dagger}_{-\sigma}d_{\sigma}(\tau)
     d^{\dagger}_{\sigma}(0)\}\rangle\rightarrow 0 \ , \label{hf12}
\end{eqnarray}
and
\begin{equation}
\langle\mathcal{T}_{\tau}\{\psi_{\alpha\bm{k}\sigma}n_{d-\sigma}(\tau)d^{\dagger}_{\sigma}(0)\}\rangle
\rightarrow -\langle n_{d-\sigma}\rangle G_{\alpha\bm{k}\sigma\sigma}(\tau) \ , \label{hf13}
\end{equation}
we find that
\begin{eqnarray}
 & & \tilde{C}_{\sigma\sigma}(i\omega_n) \label{jfeom23} \\
 & & =\frac{\langle n_{d-\sigma}\rangle}
     {i\omega_n-\epsilon_d-U+\mu+\sigma h}[1+\Sigma_0(i\omega_n)\tilde{D}_{\sigma\sigma}(i\omega_n)] \ .
     \nonumber
\end{eqnarray}
The details of the above derivation is left in appendix \ref{a2}.

Substituting Eq. (\ref{jfeom23}) into Eq. (\ref{jfeom15}), we obtain
\begin{widetext}
\begin{equation}
 \tilde{D}_{\sigma\sigma}(i\omega_n)=\frac{1}
 {i\omega_n-\epsilon_d+\mu+\sigma h-[1+\Sigma_{\sigma}(i\omega_n)]\Sigma_0(i\omega_n)} \! \left[
 1+\frac{U\langle n_{d-\sigma}\rangle}{i\omega_n-\epsilon_d-U+\mu+\sigma h}\right] , \label{jfkgf13}
 \end{equation}
\end{widetext}
where
\begin{equation}
 \Sigma_{\sigma}(i\omega_n)=\frac{U\langle n_{d-\sigma}\rangle}{i\omega_n-\epsilon_d-U+\mu+\sigma h} \ .
 \label{jfkgf14}
\end{equation}
In the $U\rightarrow+\infty$ limit, Eq. (\ref{jfkgf13}) becomes
\begin{equation}
 \tilde{D}_{\sigma\sigma}(i\omega_n)=\frac{1-\langle n_{d-\sigma}\rangle}
 {i\omega_n-\epsilon_d+\mu+\sigma h-(1-\langle n_{d-\sigma}\rangle)\Sigma_0(i\omega_n)} \ .
 \label{jfkgf15}
\end{equation}
Compared with the $U=0$ solution [Eq. (\ref{jfkgf16})], the effects of the infinite $U$ limit lie at two
aspects: (i) First of all, the impurity fermions acquire the wavefunction renormalization
$Z=\sqrt{1-\langle n_{d-\sigma}\rangle}$, which will reduce the spectral weight. (ii) Next, the
self-energy is reduced by a factor of $1-\langle n_{d-\sigma}\rangle$.

Using Eq. (\ref{jfkgf17}), we find that
\begin{eqnarray*}
 \tilde{D}_{\sigma}(\omega)=\frac{(1-\langle n_{d-\sigma}\rangle)\omega}
 {L(\omega)+\sigma h\omega+i\pi\lambda(1-\langle n_{d-\sigma}\rangle)\omega^3/D} \ ,
\end{eqnarray*}
where
$L(\omega)=[1+2\lambda(1-\langle n_{d-\sigma}\rangle)]\omega^2-\epsilon_d\omega-s\lambda D^2(1-\langle n_{d-\sigma}\rangle)$.
As a result, the spectral density is of the form
\begin{equation}
 \rho_{\sigma}(\omega)=\frac{2\pi\lambda(1-\langle n_{d-\sigma}\rangle)^2\omega^4/D}
 {[L(\omega)+\sigma h\omega]^2+[\pi\lambda(1-\langle n_{d-\sigma}\rangle)\omega^3/D]^2} \ .
 \label{jfkgf18}
\end{equation}
From Eq. (\ref{jfkgf18}), $\langle n_{d\sigma}\rangle$ are determined by the following equations
\begin{eqnarray}
 & & \langle n_{d\sigma}\rangle \label{jfchi12} \\
 & & =\! \int^{+\infty}_{-\infty} \! \! d\omega
     \frac{\lambda(1-\langle n_{d-\sigma}\rangle)^2(\omega^4/D)f(\omega)}
     {[L(\omega)+\sigma h\omega]^2+[\pi\lambda(1-\langle n_{d-\sigma}\rangle)\omega^3/D]^2} \ .
     \nonumber
\end{eqnarray}

\begin{figure}
\begin{center}
 \includegraphics[width=0.95\columnwidth]{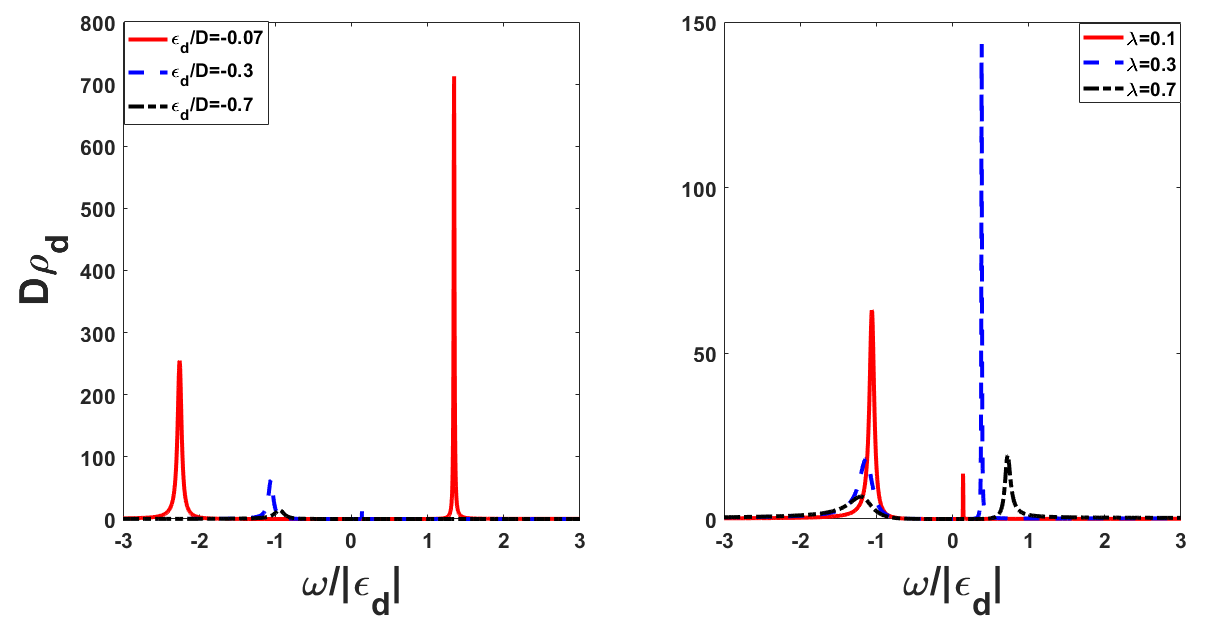}
 \caption{The spectral density $\rho_d$ in the limit $U\rightarrow+\infty$ as a function of $\omega$ (in
 units of $|\epsilon_d|$). {\bf Left}: $\lambda=0.1$ for different values of $\epsilon_d/D$. {\bf Right}:
 $\epsilon_d/D=-0.3$ for different values of $\lambda$.}
 \label{jfkondof15}
\end{center}
\end{figure}

We plot the spectral density $\rho_d$ in the limit $U\rightarrow+\infty$ as a function of $\omega$ for
various values of $\lambda$ and $\epsilon_d/D$ in Fig. \ref{jfkondof15}. When $\epsilon_d$ is close to
the Fermi energy ($\mu=0$), the spectral density has two peaks at $\omega_1<0<\omega_2$, similar to the
one at $U=0$. However, as $\epsilon_d$ moves away from the Fermi energy, the peak at $\omega_2$ is highly
suppressed compared with the one at $\omega_1$. Moreover, both $|\omega_1|$ and $\omega_2$ change from
the value about $|\epsilon_d|$ when $\epsilon_d$ is close to the Fermi energy to the value about 
$0.1|\epsilon_d|$ when $\epsilon_d$ moves away from the Fermi energy. On the other hand, for given 
$\epsilon_d$, the peak at $\omega_2$ becomes pronounced by increasing the hybridization strength 
$\lambda$. The behavior of the spectral density at different values of $\lambda$ and $\epsilon_d$ 
exhibits the competition between the hybridization and the correlation effect on the impurity level.

Two features of the spectral density should be emphasized. First of all, it has two peaks instead of a
single one as in the ordinary FL. As we have discussed before, this split of peaks arises from the flat
band. Next, the peaks in the spectral density shown in Fig. \ref{jfkondof15} are located at the
frequencies about $\pm 0.1\epsilon_d$ to $\pm\epsilon_d$, depending on the values of $\lambda$ and 
$\epsilon_d/D$. This implies the simple resonances in the infinite $U$ limit. (For the ordinary FL, the
spectral density within the same approximation will exhibit a single peak at the frequency close to 
$\epsilon_d$.) On the other hand, the peaks in the spectral density at $T=0$ (Fig. \ref{jfflf11}) move 
to the frequencies about $\pm 0.1\Delta_b$, corresponding to the Kondo resonance. This behavior in the 
spectral density clearly shows the fact that the simple resonance at high temperature ($T>T_K$) turns 
into the Kondo singlet at low temperature ($T\ll T_K$).

The above result [Eq. (\ref{jfkgf18}) and Fig. \ref{jfkondof15}] does not show the Kondo resonance. The 
situation remains similar even if we take a finite value of $U$. Thus, the approximation we have made 
does not capture the Kondo physics. The reason arises from the fact that higher-order correlations 
between the magnetic impurity and conduction electrons are neglected within this approximation [Eqs. 
(\ref{hf12}) and (\ref{hf13})], in particular the spin-flip processes. These processes become important 
at low temperatures, i.e., $T<T_K$, and give rise to Kondo screening. However, it does include the 
correlation brought about by the infinite $U$, as shown by the following temperature dependence of $n_d$. 
Similar situations are encountered in the study of Kondo effect in a FL\cite{Varma} and charge transport 
through a quantum dot due to Coulomb blockade\cite{Meir}. Therefore, we expect that the Hartree-Fock 
approximation provides a good description on the physics at $T>T_K$.

When $h=0$, $\langle n_{d+}\rangle=\langle n_{d-}\rangle=n_d/2$, and Eq. (\ref{jfchi12}) reduces to
\begin{equation}
 \frac{n_d}{2}=\! \int^{+\infty}_{-\infty} \! \! d\omega\frac{\lambda(1-n_d/2)^2(\omega^4/D)f(\omega)}
 {[L(\omega)]^2+[\pi\lambda(1-n_d/2)\omega^3/D]^2} \ . \label{jfchi13}
\end{equation}
The value of $n_d$ can be determined by solving Eq. (\ref{jfchi13}). To obtain $\chi_{imp}$, we write
$\langle n_{d\sigma}\rangle=n_d/2+\gamma\sigma h/D+O(h^2)$ where $\gamma$ is a constant independent of
$h$. Then, we have $\chi_{imp}=2\gamma/D$. Inserting this expansion into Eq. (\ref{jfchi12}), we have
$\gamma=A/(1-B)$, where
\begin{eqnarray}
 A \! \! &=& \! \! -\! \int^{+\infty}_{-\infty} \! \! d\omega
 \frac{2\lambda(1-n_d/2)^2\omega^5L(\omega)f(\omega)}
 {\{[L(\omega)]^2+[\pi\lambda(1-n_d/2)\omega^3/D]^2\}^2} , \nonumber \\
 B \! \! &=& \! \! \! \int^{+\infty}_{-\infty} \! \! d\omega
 \frac{2\lambda(1-n_d/2)(\omega^5/D)(\omega-\epsilon_d)L(\omega)f(\omega)}
 {\{[L(\omega)]^2+[\pi\lambda(1-n_d/2)\omega^3/D]^2\}^2} . ~~~~~~\label{jfchi14}
\end{eqnarray}
Substituting the value of $n_d$ into Eq. (\ref{jfchi14}) and then performing the integrals, we get
$\chi_{imp}(T)$. On account of the approximation we have made, the resulting single-particle Green
function for impurity fermions captures the correlation in the $U\rightarrow+\infty$, but fails to
produce the Kondo physics. Thus, the above results can be applied only to the high temperature regime,
i.e., $T>T_K$, where $T_K=O(\Delta_b)$ is the Kondo temperature.

\begin{figure}
\begin{center}
 \includegraphics[width=0.95\columnwidth]{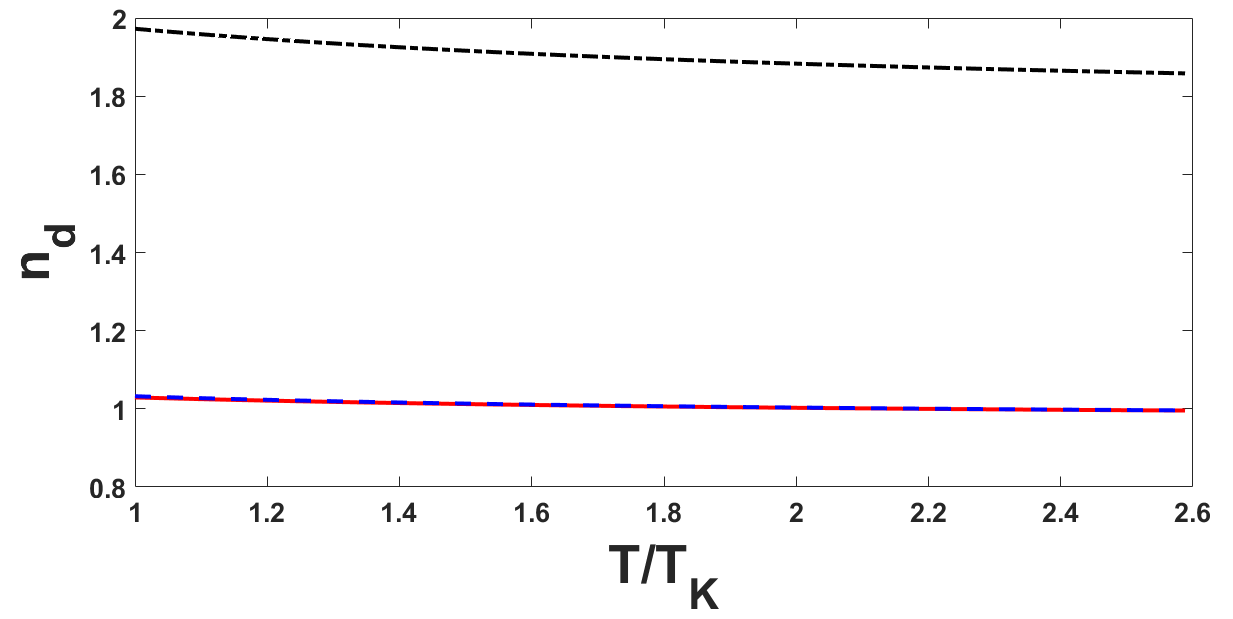}
 \caption{The occupation number $n_d$ of the impurity level as a function of $T$ (for $T>T_K$) with
 $\lambda=0.07$ and $\epsilon_d/D=-0.1$ for the $J=1$ fermions at an infinite $U$ (the solid line), the
 flat band being excluded at an infinite $U$ (the dashed line), and the $J=1$ fermions at $U=0$ (the
 dashed-dotted line). For the given parameters, $T_K/D=0.1932$.}
 \label{jfkondof13}
\end{center}
\end{figure}

We plot $n_d$ as a function of temperature $T$ for given values of $\epsilon_d/D$ and $\lambda$ at $U=0$
and $U\rightarrow+\infty$ in Fig. \ref{jfkondof13}. We see that $n_d$ at an infinite $U$ is much smaller
than that at $U=0$, reflecting the correlation brought about by the infinite $U$. Moreover, $n_d$ is a
monotonically decreasing function of $T$ when $T>T_K$, which is similar to the case in the FL\cite{Varma}.
The artificial case with the flat band being excluded is also sketched for comparison. When $T>T_K$,
i.e., the temperature range in which the approximation holds, both have the same value, indicating the
minor role played by the flat band at high temperatures.

\begin{figure}
\begin{center}
 \includegraphics[width=0.95\columnwidth]{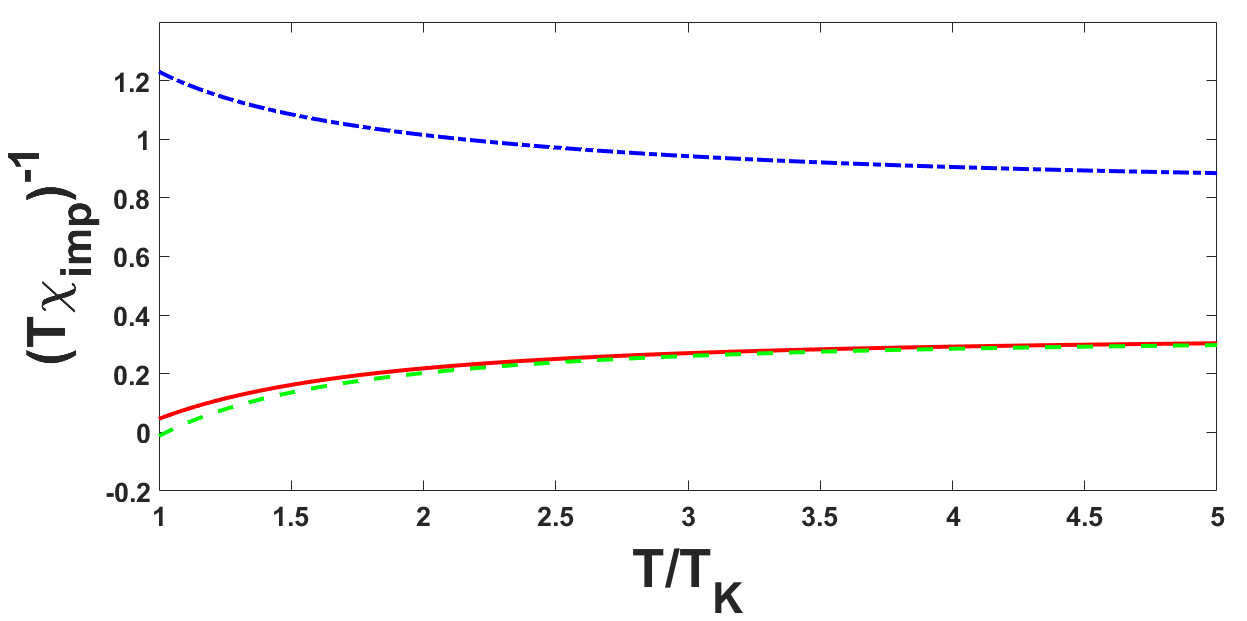}
 \caption{$(T\chi_{imp})^{-1}$ as a function of $T$ (for $T>T_K$) with $\lambda=0.07$ and
 $\epsilon_d/D=-0.1$ for the $J=1$ fermions at an infinite $U$ (the solid line), the flat band being
 excluded at an infinite $U$ (the dashed line), and the $J=1$ fermions at $U=0$ (the dashed-doted line).
 For the given parameters, $T_K/D=0.1932$.}
 \label{jfkondof14}
\end{center}
\end{figure}

The temperature dependence of $\chi_{imp}$ at $T>T_K$ for given values of $\epsilon_d/D$ and $\lambda$
at $U=0$ and $U\rightarrow+\infty$ is sketched in Fig. \ref{jfkondof14}. We see that $\chi_{imp}$
exhibits a Curie form at $T\gg T_K$ and the deviation from it occurs when $T$ is close to $T_K$, similar
to the case in a FL. This must be the case since the impurity behaves like a free moment at high
temperatures in the local moment regime, while Kondo effect starts to function near $T_K$. The
correlation brought about by an infinite $U$ increases the value of $\chi_{imp}$ compared with the one
at $U=0$. Moreover, the trends of $\chi_{imp}$ for $U=0$ and infinite $U$ are opposite when $T$ is close
to $T_K$. When the flat band is excluded by setting $s=0$, the temperature dependence of $\chi_{imp}$ at
$T\gg T_K$ in similar to that in the presence of the flat band. The distinction is visible only close to
the Kondo temperature. There, $\chi_{imp}$ becomes a non-analytic function of $T$ in the absence of the
flat band since $\chi_{imp}^{-1}$ crosses zero and becomes negative.

\section{Conclusions}

In the present work, we study the Kondo physics of a single magnetic impurity in a $J=1$ fermion system
by analyzing the Anderson impurity model in the infinite $U$ limit. As we have argued in the introduction,
the perturbative RG cannot capture the low-temperature physics in this situation due to the presence of 
a flat band. Hence, we employ two methods -- the variational wavefunction and the EOM to examine the
physical properties of this system. These two approaches are, in fact, complementary to each other.

Actually, there are two types of three-band touching fermion systems in the literature. The first type
has a strong SOC\cite{Bradlyn}, while the SOC is very weak in the second type\cite{Manes,PTang,GChang}. 
The latter is expected to be realized in the CoSi family and has been confirmed by the ARPES\cite{exp}. 
For the second type, the electron spin plays a role similar to that in graphene. The model we have 
studied in the present work describes this situation. Hence, we expect that some of our results can be 
observed in the CoSi family.

With the help of a mean-field approximation, we can determine the phase diagram of the Anderson impurity
model. Similar to the magnetic impurity in a FL, there are two regimes -- the simple resonance and the
local moment regime, due to the competition between hybridization and the on-site Coulomb repulsion $U$.
In the simple resonance regime, the hybridization between impurity and conduction electrons turns the
impurity level to a virtual bound state and there is no local moment. The properties in this regime can
be described qualitatively by the solution at $U=0$. On the other hand, the Kondo effect occurs only in
the local moment regime. In contrast with the FL, the local moment regime in the $J=1$ fermions can be
extended to the $\epsilon_d>0$ region when the value of $U$ is large enough. As we have discussed, this
is related to the existence of the flat band. When the temperature is below the Kondo temperature $T_K$,
Kondo effect occurs, as we have shown in the approach of variational wavefunction, and the local moments
are screened by conduction electrons.

In terms of the variational wavefunction approach, we can calculate the binding energy and show that the
Kondo screening always occurs as long as the exchange coupling between the impurity spin and the
conduction electrons is antiferromagnetic in nature, similar to the case in the FL. We also calculate 
the impurity contribution to the magnetic susceptibility and the local electron occupation number at the
impurity site at $T=0$, which depend on the DOS of the flat band in a nontrivial way.

Following the previous local-FL description of the Kondo effect\cite{Hewson}, we employ an effective
Hamiltonian describing the low-temperature physics in the local moment regime in the infinite $U$ limit.
In terms of the above calculated ground-state properties, we can relate the parameters in the effective
Hamiltonian to those in the infinite-$U$ Anderson model. Consequently, we are able to determine the
impurity spectral density at $T=0$ and the impurity contribution to the heat capacity at low temperature.
Especially, we show that the resulting Kondo resonance in this system is split into two peaks due to the
presence of the flat band.

The physics in the local moment regime can be qualitatively described by the infinite $U$ limit. We then
use the method of EOM to calculate the single-particle Green function of the impurity fermions in this
limit, from which we can extract the temperature dependence of the occupation number $n_d$ of impurity
fermions as well as the impurity magnetic susceptibility. Both are similar to those in a FL, as we would
expect. Especially, $\chi_{imp}$ exhibits a Curie-like behavior at $T\gg T_K$ and is much enhanced near
$T_K$. On the other hand, the flat band has no effect on the temperature dependence of $n_d$ as long as
$T>T_K$, so that the latter is identical to the case in a WSM. The role of the flat band reveals itself
only when the temperature is close to $T_K$. If the flat band were absent, $\chi_{imp}$ would become
non-analytic function of $T$ near $T_K$.

When $U\neq 0$, the set of EOMs cannot be solved exactly and we make the Hartree-Fock approximation, 
which takes into account the correlation brought about by an infinite $U$ but misses the Kondo resonance. 
Therefore, our results on the temperature dependence of the occupation number of impurity fermions and 
the impurity magnetic susceptibility can be applied only to the temperature regime $T>T_K$. Actually, a 
more sophisticated approximation can be made to capture the Kondo physics\cite{Lacroix}. Also, the 
approaches adopted in the present work can be directly applied to other multiband touching fermion 
systems and can be used to study the anisotropic correlations introduced by the velocity anisotropy 
and/or the tilting of the dispersion. These will be left as future works.

For simplicity, we have assumed that the hybridization between the impurity and the three bands have the
same strength. We expect that small deviation from this isotropic limit will not affect the physics we
have described qualitatively. Especially, the Kondo screening always occurs in the local moment regime
as long as the impurity couples to the flat band. Nevertheless, in the highly anisotropic limit, i.e.,
the strength of the coupling to the flat band is much smaller than those to the topologically nontrivial
bands, the resulting Kondo temperature may be too small to be accessible by experiments. In this
situation, experimental data may suggest the pseudogap Kondo effect.

In the present work, we have neglected the electron-electron interaction between conduction electrons.
That is, we have assumed that there is a window for the interaction strength such that the three-band
touching point is stable. Since the band structure of CoSi near the $\Gamma$ point observed by the
ARPES and that obtained by the {\it ab initio} calculations are both qualitatively consistent with the 
one of non-interacting $J=1$ fermions, this assumption is at least valid in the CoSi family.

In the study of the Kondo physics in graphene\cite{Fritz2}, due to the presence of two valleys or Dirac 
nodes in the Brillouin zone, the issue of whether or not the two-channel Kondo physics is relevant at 
low temperature was raised. The is because electrons from the two valleys form two independent screening
channels at low energy. For the $J=1$ fermions with two nodes at the Fermi energy, the relevancy of the 
two-channel Kondo effect is an interesting open problem. However, for the CoSi family we studied in this 
paper, band structure calculations and the ARPES show that there is only a single nodal point at a given 
energy. Therefore, such an issue does not exist. In the present work, we consider only the simplest 
spin-$1/2$ impurity. When the impurity spin is larger than one-half, the underscreened Kondo effect may 
occur. All the above interesting questions will be left for future studies.

\acknowledgments

The works of Y.L. Lee and Y.-W. Lee are supported by the Ministry of Science and Technology, Taiwan,
under the grant number MOST 108-2112-M-018-005 and MOST 108-2112-M-029-002, respectively.

\appendix
\section{Derivation of Eq. (\ref{jfgse15})}
\label{a1}

Here we present the derivation of Eq. (\ref{jfgse15}). From the conditions
\begin{eqnarray*}
 \frac{\partial E}{\partial a_0^*}=0=\frac{\partial E}{\partial a_{\alpha\bm{k}\sigma}^*} \ ,
\end{eqnarray*}
we get
\begin{equation}
 Ea_0= \! \sum_{\bm{k},\sigma,\alpha}^{\prime} \! \left[(\alpha vk-\mu)a_0+\frac{1}{\sqrt{\Omega}}
 \tilde{V}_{\bm{k}\alpha}a_{\alpha\bm{k}\sigma}\right] , \label{jfgse1}
\end{equation}
and
\begin{equation}
 [\alpha vk-\mu-\Delta_b-(\eta_h-\sigma)h]a_{\alpha\bm{k}\sigma}=\frac{1}{\sqrt{\Omega}}
 \tilde{V}^*_{\bm{k}\alpha}a_0 \ , \label{jfgse11}
\end{equation}
where $\Delta_b\equiv E_0-E$ is the binding energy and $\eta_h=\mbox{sgn}(h)$. Using Eq. (\ref{jfgse11})
to eliminate $a_{\alpha\bm{k}\sigma}$ and noticing that $a_0\neq 0$, we get
\begin{equation}
 \epsilon_d-\mu-|h|-\Delta_b=\frac{1}{\Omega} \! \sum_{\bm{k},\sigma,\alpha}^{\prime}
 \frac{|\tilde{V}_{\bm{k}\alpha}|^2}{\alpha vk-\mu-\Delta_b-(\eta_h-\sigma)h} \ . \label{jfgse12}
\end{equation}

To simplify Eq. (\ref{jfgse12}), we define the $3\times 3$ matrix $\hat{G}(z,\bm{k})$ whose elements are
given by
\begin{equation}
 G_{\alpha\beta}(z,\bm{k})\equiv\frac{1}{z-\alpha vk}\delta_{\alpha\beta} \ . \label{jfgse13}
\end{equation}
Then, Eq. (\ref{jfgse12}) can be written as
\begin{equation}
 \epsilon_d-\mu-|h|-\Delta_b=-\frac{|V|^2}{\Omega} \! \sum_{\bm{k},\sigma}^{\prime} \! \!
 \sum_{\alpha,\beta} \! \! \left[U^{\dagger}(\bm{k})G(z,\bm{k})U(\bm{k})\right]_{\alpha\beta} ,
 \label{jfgse14}
\end{equation}
where $z=\mu+\Delta_b+(\eta_h-\sigma)h$. With the help of Eq. (\ref{jfkh25}), we find that
\begin{eqnarray*}
 \! \sum_{\bm{k}}^{\prime}U^{\dagger}(\bm{k})G(z,\bm{k})U(\bm{k})=\frac{1}{3}I \! \sum_{\bm{k}}^{\prime}
 \! \sum_{\alpha}\frac{1}{z-\alpha vk} \ ,
\end{eqnarray*}
where $I$ is the $3\times 3$ unit matrix. Since $\sum_{\alpha\beta}I_{\alpha\beta}=3$, Eq. (\ref{jfgse14})
becomes Eq. (\ref{jfgse15}).

\section{Derivation of the impurity Green function}
\setcounter{equation}{0}
\label{a2}

Following the standard procedure, the EOM of $D_{\sigma\sigma^{\prime}}(\tau)$ is given by
\begin{eqnarray}
 & & \! \! \! \! \! \! -\partial_{\tau}D_{\sigma\sigma^{\prime}}(\tau)=\delta(\tau)
     \delta_{\sigma\sigma^{\prime}}+(\epsilon_d-\mu-\sigma h)D_{\sigma\sigma^{\prime}}(\tau) \nonumber
     \\
 & & ~~~+\frac{1}{\sqrt{\Omega}} \! \sum_{\bm{k},\alpha}\tilde{V}^*_{\bm{k}\alpha}
     G_{\alpha\bm{k}\sigma\sigma^{\prime}}(\tau)+UC_{\sigma\sigma^{\prime}}(\tau) \ , \label{jfeom1}
\end{eqnarray}
where $C_{\sigma\sigma^{\prime}}(\tau)$ is defined as Eq. (\ref{jfkgf11}) and
\begin{equation}
 G_{\alpha\bm{k}\sigma\sigma^{\prime}}(\tau)\equiv -\langle\mathcal{T}_{\tau}\{
 \psi_{\alpha\bm{k}\sigma}(\tau)d^{\dagger}_{\sigma^{\prime}}(0)\}\rangle \ . \label{jfkgf110}
\end{equation}
By taking the Fourier transform on both sides of Eq. (\ref{jfeom1}), we get
\begin{eqnarray}
 & & \! \! \! \! i\omega_n\tilde{D}_{\sigma\sigma^{\prime}}(i\omega_n)=\delta_{\sigma\sigma^{\prime}}
     +(\epsilon_d-\mu-\sigma h)\tilde{D}_{\sigma\sigma^{\prime}}(i\omega_n) \nonumber \\
 & & ~~+\frac{1}{\sqrt{\Omega}} \! \sum_{\bm{k}\alpha}\tilde{V}^*_{\bm{k}\alpha}
     \tilde{G}_{\alpha\bm{k}\sigma\sigma^{\prime}}(i\omega_n)+U\tilde{C}_{\sigma\sigma^{\prime}}(i\omega_n)
     \ , ~~~~\label{jfeom12}
\end{eqnarray}
where $\omega_n=(2n+1)\pi T$ and $\tilde{G}(i\omega_n)$ is the Fourier transform of $G(\tau)$.

Similarly, the EOM of $G_{\alpha\bm{k}\sigma\sigma^{\prime}}(\tau)$ is given by
\begin{equation}
 -\partial_{\tau}G_{\alpha\bm{k}\sigma\sigma^{\prime}}(\tau)=(\alpha vk-\mu)
 G_{\alpha\bm{k}\sigma\sigma^{\prime}}(\tau)+\frac{\tilde{V}_{\bm{k}\alpha}}{\sqrt{\Omega}}
 D_{\sigma\sigma^{\prime}}(\tau) \ . \label{jfeom13}
\end{equation}
By taking the Fourier transform on both sides of Eq. (\ref{jfeom13}), we get
\begin{equation}
 \tilde{G}_{\alpha\bm{k}\sigma\sigma^{\prime}}(i\omega_n)=\frac{1}{\sqrt{\Omega}}
 \frac{\tilde{V}_{\bm{k}\alpha}}{i\omega_n-\alpha vk+\mu}\tilde{D}_{\sigma\sigma^{\prime}}(i\omega_n) \ .
 \label{jfeom14}
\end{equation}
Using Eq. (\ref{jfeom14}) to eliminate $\tilde{G}_{\alpha\bm{k}\sigma\sigma^{\prime}}(i\omega_n)$, we
obtain Eq. (\ref{jfeom15}).

When $U\neq 0$, we need the EOM of $C_{\sigma\sigma}(\tau)$, which is given by
\begin{widetext}
\begin{eqnarray}
	-\partial_{\tau}C_{\sigma\sigma}(\tau) &=& \delta(\tau)\langle n_{d-\sigma}\rangle
	+(\epsilon_d-\mu-\sigma h+U)C_{\sigma\sigma}(\tau)-\frac{1}{\sqrt{\Omega}} \! \sum_{\bm{k},\alpha}
	\tilde{V}^*_{\bm{k}\alpha}\langle\mathcal{T}_{\tau}\{\psi_{\alpha\bm{k}\sigma}n_{d-\sigma}(\tau)
	d^{\dagger}_{\sigma}(0)\}\rangle \nonumber \\
	& & +\frac{1}{\sqrt{\Omega}} \! \sum_{\bm{k},\alpha}\tilde{V}_{\bm{k}\alpha}\langle\mathcal{T}_{\tau}\{
	\psi^{\dagger}_{\alpha\bm{k}-\sigma}d_{-\sigma}d_{\sigma}(\tau)d^{\dagger}_{\sigma}(0)\}\rangle+\frac{1}
	{\sqrt{\Omega}} \! \sum_{\bm{k},\alpha}\tilde{V}^*_{\bm{k}\alpha}\langle\mathcal{T}_{\tau}\{
	\psi_{\alpha\bm{k}-\sigma}d^{\dagger}_{-\sigma}d_{\sigma}(\tau)d^{\dagger}_{\sigma}(0)\}\rangle \ .
	\label{jfeom2}
\end{eqnarray}
In the above, we have used the identity $n_{d\sigma}^2=n_{d\sigma}$. To proceed, we have to make an
approximation to obtain a closed set of EOMs. Within the Hartree-Fock approximation [Eqs. (\ref{hf12})
and (\ref{hf13})], Eq. (\ref{jfeom2}) can be approximated as
\begin{equation}
	-\partial_{\tau}C_{\sigma\sigma}(\tau)\approx\delta(\tau)\langle n_{d-\sigma}\rangle
	+(\epsilon_d-\mu-\sigma h+U)C_{\sigma\sigma}(\tau)+\frac{1}{\sqrt{\Omega}} \! \sum_{\bm{k},\alpha}
	\tilde{V}^*_{\bm{k}\alpha}\langle n_{d-\sigma}\rangle G_{\alpha\bm{k}\sigma\sigma}(\tau) \ .
	\label{jfeom21}
\end{equation}
By taking the Fourier transform on both sides of Eq. (\ref{jfeom21}), we find that
\begin{equation}
	\tilde{C}_{\sigma\sigma}(i\omega_n)=\frac{\langle n_{d-\sigma}\rangle}
	{i\omega_n-\epsilon_d-U+\mu+\sigma h} \! \left[1+\frac{1}{\sqrt{\Omega}}\! \sum_{\bm{k},\alpha}
	\tilde{V}^*_{\bm{k}\alpha}\tilde{G}_{\alpha\bm{k}\sigma\sigma}(i\omega_n)\right] . \label{jfeom22}
\end{equation}
\end{widetext}
With the help of Eq. (\ref{jfeom14}), we get Eq. (\ref{jfeom23}).

\section{The $U=0$ solution}
\label{resonance}
\label{a3}

Although our main interest is to study the Kondo physics by analyzing the Anderson model in the infinite
$U$ limit, the $U=0$ solution presented below can be used as a benchmark for a comparison between the
physics in the local moment regime and that in the simple resonance regime.

When $U=0$, $\tilde{D}_{\sigma\sigma^{\prime}}(i\omega_n)$ can be exactly determined from Eq.
(\ref{jfeom15}), yielding
\begin{equation}
\tilde{D}_{\sigma\sigma^{\prime}}(i\omega_n)=\frac{\delta_{\sigma\sigma^{\prime}}}
{i\omega_n-\epsilon_d+\sigma h-\Sigma_0(i\omega_n)} \ . \label{jfkgf16}
\end{equation}
The retarded self-energy $\Sigma_{R0}(\omega)$ is given by
\begin{eqnarray}
& & \Sigma_{R0}(\omega)=\! \int^{+\infty}_{-\infty} \! \! d\epsilon\frac{|V|^2N(\epsilon)}
{\omega-\epsilon+i0^+} \nonumber \\
& & =\frac{s\lambda D^2}{\omega}-2\lambda\omega-i\pi s\lambda D^2\delta(\omega)-i\frac{\pi\lambda}{D}
\omega^2 \ , \label{jfkgf17}
\end{eqnarray}
where $s=A/(CD^3)=1/3$. Thus, the retarded single-particle Green function for the impurity fermions is
of the form
$\tilde{D}_{R\sigma\sigma^{\prime}}(\omega)=\tilde{D}_{\sigma}(\omega)\delta_{\sigma\sigma^{\prime}}$,
where
\begin{equation}
\tilde{D}_{\sigma}(\omega)=\frac{\omega}{L_0(\omega)+\sigma h\omega+i\pi\lambda\omega^3/D} \ ,
\label{jfkgf10}
\end{equation}
where $L_0(\omega)=(1+2\lambda)\omega^2-\epsilon_d\omega-s\lambda D^2$. The spectral density
$\rho_{\sigma}(\omega)=-2\mbox{Im}\tilde{D}_{\sigma}(\omega)$ is then given by
\begin{equation}
\rho_{\sigma}(\omega)=\frac{2\pi\lambda\omega^4/D}{[L_0(\omega)+\sigma h\omega]^2+(\pi\lambda\omega^3/D)^2}
\ . \label{jfkgf12}
\end{equation}

\begin{figure}
	\begin{center}
		\includegraphics[width=0.95\columnwidth]{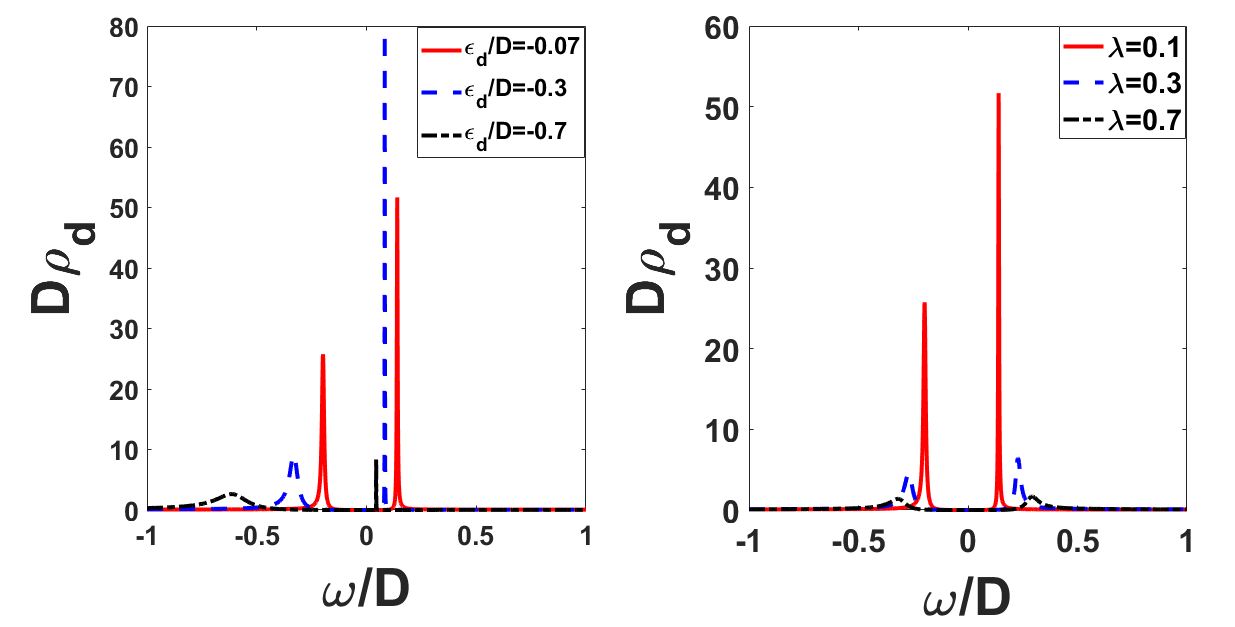}
		\caption{The spectral density $\rho_d$ at $U=0$ as a function of $\omega$. {\bf Left}: $\lambda=0.1$
			for different values of $\epsilon_d/D<0$. {\bf Right}: $\epsilon_d/D=-0.3$ for different values of
			$\lambda$.}
		\label{jfkondof12}
	\end{center}
\end{figure}

Figure \ref{jfkondof12} shows the spectral density $\rho_d=\rho_{\sigma}|_{h=0}$ as a function of $\omega$
for different values of $\epsilon_d/D$ and $\lambda$. We see that $\rho_d$ exhibits two peaks at
$\omega=\omega_{\pm}$ where $\omega_-<\omega_+$. When $\lambda\ll 1$, the positions of the peaks can be
estimated by the zeros of the function $L_0(\omega)$, i.e., the solutions of the equation $L_0(\omega)=0$,
leading to
\begin{eqnarray*}
 \omega_{\pm}\approx\frac{\epsilon_d\pm\sqrt{\epsilon_d^2+4s\lambda D^2(1+2\lambda)}}{2(1+2\lambda)} \ .
\end{eqnarray*}
For given $\lambda$, both $\omega_{\pm}$ will shift to smaller values as $|\epsilon_d/D|$ increases.
Moreover, the peak values of $\rho_d$ will decrease with increasing $|\epsilon_d/D|$. On the other hand,
for given $\epsilon_d/D<0$, $\omega_-$ will move to a smaller value while $\omega_+$ will move to a
larger value as $\lambda$ increases. The peak values of $\rho_d$ decrease with increasing $\lambda$.

For the magnetic impurity in a FL, the spectral density at $U=0$ exhibits a peak at the impurity level,
$\epsilon_d$. The only effect of the hybridization is to broaden this peak and turns the impurity level
into a virtual bound state or resonance. For the case in the $J=1$ fermions, the existence of the flat
band leads to a two-peak structure in the spectral density and shifts them away from the position of the
impurity level. On the other hand, the nonzero width of these peaks arises mainly from the hybridization
with the topologically nontrivial bands.

The total occupation number $n_d$ of the impurity levels at $h=0$ is given by
\begin{equation}
 n_d=\! \int^{+\infty}_{-\infty} \! \! d\omega\frac{(2\lambda\omega^4/D)f(\omega)}
 {[L_0(\omega)]^2+(\pi\lambda\omega^3/D)^2} , \label{jfchi1}
\end{equation}
where $f(\omega)=1/(e^{\beta\omega}+1)$ is the Fermi-Dirac distribution function. The impurity
contribution to the magnetic susceptibility is then of the form
\begin{equation}
 \chi_{imp}(T)=- \! \int^{+\infty}_{-\infty} \! \! d\omega\frac{(8\lambda\omega^5/D)L_0(\omega)f(\omega)}
 {\{[L_0(\omega)]^2+(\pi\lambda\omega^3/D)^2\}^2} . \label{jfchi11}
\end{equation}
The $U=0$ solution is supposed to capture the physical properties of the simple resonance regime
qualitatively.



\begin{thebibliography}{99}
\bibitem{foot1} This can be seen by introducing the dimensionless coupling $\lambda\equiv C\Lambda^2K$.
        Then, from Eq. (\ref{jfkrg1}), the one-loop RG equation for $\lambda$ is
        \begin{eqnarray*}
         \frac{d\lambda}{dl}=-2\lambda+\frac{\lambda^2}{2} \ .
        \end{eqnarray*}
        This equation has two fixed points: $\lambda=0$ and $\lambda=\lambda_c=4$. For the
        antiferromagnetic coupling ($\lambda>0$), the former is IR stable while the latter is IR
        unstable. Thus, for $\lambda<\lambda_c$, $\lambda$ flows to zero at low energies, implying
        decoupled impurity spin and conduction electrons. On the other hand, $\lambda$ flows to strong
        coupling at low energies when $\lambda>\lambda_c$, implying the occurrence of Kondo screening.
        As a result, the Kondo screening occurs only when $K$ is larger than some critical strength in
        this case.
\end{thebibliography}
\end{document}